\documentclass[apj]{emulateapj}
\usepackage{graphicx}
\usepackage{natbib}
\usepackage{rotating}
\usepackage{adjustbox}

\shorttitle{Abundances of PNe in the substructures of M31}
\shortauthors{X.\ Fang et al.}

\begin{document}

\title{Chemical Abundances of Planetary Nebulae in the Substructures of 
M31\footnotemark[$\ast$]}

\footnotetext[$\ast$]{Based on observations made with the Gran Telescopio 
Canarias, installed at the Spanish Observatorio del Roque de los Muchachos 
of the Instituto de Astrof\'{i}sica de Canarias, in the island of La Palma.
These observations are associated with program No. GTC55-14B.
}

\author{
Xuan Fang$^{1}$, 
Rub\'{e}n Garc\'{i}a-Benito$^{1}$,
Mart\'{i}n A.\ Guerrero$^{1}$, 
Xiaowei Liu$^{2, 3}$, 
Haibo Yuan$^{3, 4}$, \\
Yong Zhang$^{5}$,
and
Bing Zhang$^{2}$\\
}
\affil{
$^{1}$Instituto de Astrof\'{i}sica de Andaluc\'{i}a (IAA-CSIC), Glorieta 
de la Astronom\'{i}a s/n, E-18008 Granada, Spain\\
$^{2}$Department of Astronomy, School of Physics, Peking University, Beijing 100871, China\\
$^{3}$Kavli Institute for Astronomy and Astrophysics, Peking University, Beijing 100871, China\\
$^{4}$LAMOST Fellow\\
$^{5}$Department of Physics, University of Hong Kong, Pokfulam Road, Hong Kong, China
}
\email{fangx@iaa.es}

\begin{abstract}

We present deep spectroscopy of planetary nebulae (PNe) that are associated 
with the substructures of the Andromeda Galaxy (M31).  The spectra were 
obtained with the OSIRIS spectrograph on the 10.4\,m GTC.  Seven targets 
were selected for the observations, three in the Northern Spur and four 
associated with the Giant Stream.  The most distant target in our sample, 
with a rectified galactocentric distance $\geq$100~kpc, was the first PN 
discovered in the outer streams of M31.  The [O~{\sc iii}] $\lambda$4363 
auroral line is well detected in the spectra of all targets, enabling 
electron temperature determination.  Ionic abundances are derived based 
on the [O~{\sc iii}] temperatures, and elemental abundances of helium, 
nitrogen, oxygen, neon, sulfur, and argon are estimated.  The relatively 
low N/O and He/H ratios as well as abundance ratios of $\alpha$-elements 
indicate that our target PNe might belong to populations as old as 
$\sim$2~Gyr.  Our PN sample, including the current seven and the previous 
three observed by Fang et al., have rather homogeneous oxygen abundances. 
The study of abundances and the spatial and kinematical properties 
of our sample leads to the tempting conclusion that their progenitors 
might belong to the same stellar population, which hints at a possibility 
that the Northern Spur and the Giant Stream have the same origin.  This 
may be explained by the stellar orbit proposed by Merrett et al. 
Judging from the position and kinematics, we emphasize that M32 might be 
responsible for the two substructures.  Deep spectroscopy of PNe in M32 
will help to assess this hypothesis. 

\end{abstract}


\section{Introduction} \label{section1}

Merging of smaller galaxies makes major contributions to the growth 
of a large galaxy \citep[e.g.,][]{whi78,wr78}.  The relics of such 
interaction can be preserved in the outskirts of galaxies and detected as 
various substructures such as tidal tails or stellar streams 
\citep[e.g.,][]{iba01a,maj03}.  The Andromeda Galaxy (M31), being the 
nearest large spiral system, is probably the best candidate for studying 
the debris of galaxy interaction/formation.  Numerous stellar substructures 
have been detected in its outskirts \citep[e.g.,][]{iba01b,iba07,fer02,
irw05,mccon03,mccon04,mccon09}.  The morphologies and stellar populations 
of the substructures provide important clues to the assemblage history of 
the galaxy.  However, due to the intrinsic faintness and the vast spatial 
extension of these substructures, a comprehensive survey of the stars 
therein is difficult.  As descendants of the low- and intermediate-mass 
stars, planetary nebulae (PNe) are ubiquitous in the universe and 
excellent tracers to study the chemistry, kinematics and stellar 
contents of the substructures.  Being bright in narrow emission lines, 
PNe are easily detectable at the distance of M31 
\citep[785~kpc;][]{mccon05}, and their spectra can be used to measure 
radial velocities and derive accurate abundances of elements such as He, 
O, C, N, Ne, Ar and S.

Among all substructures found in M31, the southern stellar stream 
(also known as the Giant Stream; \citealt{iba01b,cald10}) and the 
Northern Spur are the most prominent.  The Northern Spur was originally 
observed to lie in the direction of M31's 
gaseous warp \citep[][]{ne77}, and \citet{fer02} found enhancement in both 
stellar density and metallicity in the Northern Spur through a panoramic 
survey of the red giant branch (RGB) stars in M31.  The Giant Stream was 
discovered by \citet{iba01b} as a stream of metal-rich stars within the 
halo of M31.  \citet{fer02} proposed the possible connection between 
the Giant Stream and the Northern Spur.  \citet{mer03} presented a 
three-dimensional orbit for the stellar streams that connects the two 
substructures by studying the kinematics of $\sim$20 PNe in the disk of 
M31.  It has been suggested that dwarf galaxies such as M32 and NGC\,205 
might be responsible for the two substructures \citep[e.g.,][]{iba01b,mer03} 
mainly because the two satellites are spatially associated with the Stream.

Today the exact origin of the Giant Stream and the Northern Spur of M31 
is still unclear, and observational data for them are scarce. 
Spectroscopy of the PNe in these two substructures can provide accurate 
chemical information to shed light on this problem.  With this purpose, 
\citet[][hereafter Paper~I]{fang13} 
carried out spectroscopic observations of three PNe in the Northern Spur 
using the Double Spectrograph on the 5.1\,m Hale telescope at Palomar 
Observatory.  The authors found relatively high oxygen abundance in one 
of the three targets compared to the disk sample of \citet{kwi12} at 
similar galactocentric distance, a result that might be in line with 
the paradigm of tidal disruption of M31 satellites. 
\citet{balick13} observed two halo/outer-disk PNe in M31 using the 10.4\,m 
Gran Telescopio Canarias (GTC).  The authors invoked starburst triggered by 
M31--M33 encounter to explain the unexpected high oxygen abundances found 
in these two PNe.  More recent observations of outskirt PNe by 
\citet{cor15} support the current view that the external regions of M31 
are the result of complex interactions and merger processes.

We have carried out deep spectroscopy of seven PNe in the Northern Spur 
and the Giant Stream of M31 using the 10.4\,m GTC.  The purpose of these 
observations is to derive reliable elemental abundances of He, O, N, Ne,
Ar and S.  We expect excellent detection of the temperature-sensitive 
auroral lines of heavy elements (i.e., [O~{\sc iii}] $\lambda$4363, and 
the even fainter [N~{\sc ii}] $\lambda$5755 line), given the high 
light-collection efficiency of GTC and excellent observing conditions 
at Observatorio de Roque de los Muchachos (ORM). 
This paper is a follow-up project of Paper~I to investigate the 
origin of M31 substructures by studying the chemistry of PNe therein. 
Section~\ref{section2} presents the observations and data reduction. 
Line fluxes, plasma diagnostics and abundance determinations are 
given in Section~\ref{section3}.  We discuss the spatial distribution 
and kinematics, abundance correlations, stellar population of the PN 
progenitors, the radial distribution of oxygen, and the origin of the 
substructures in Section~\ref{section4}.  Finally, we make a summary 
and present conclusions in Section~\ref{section5}.

\begin{figure}
\begin{center}
\includegraphics[width=1.0\columnwidth,angle=0]{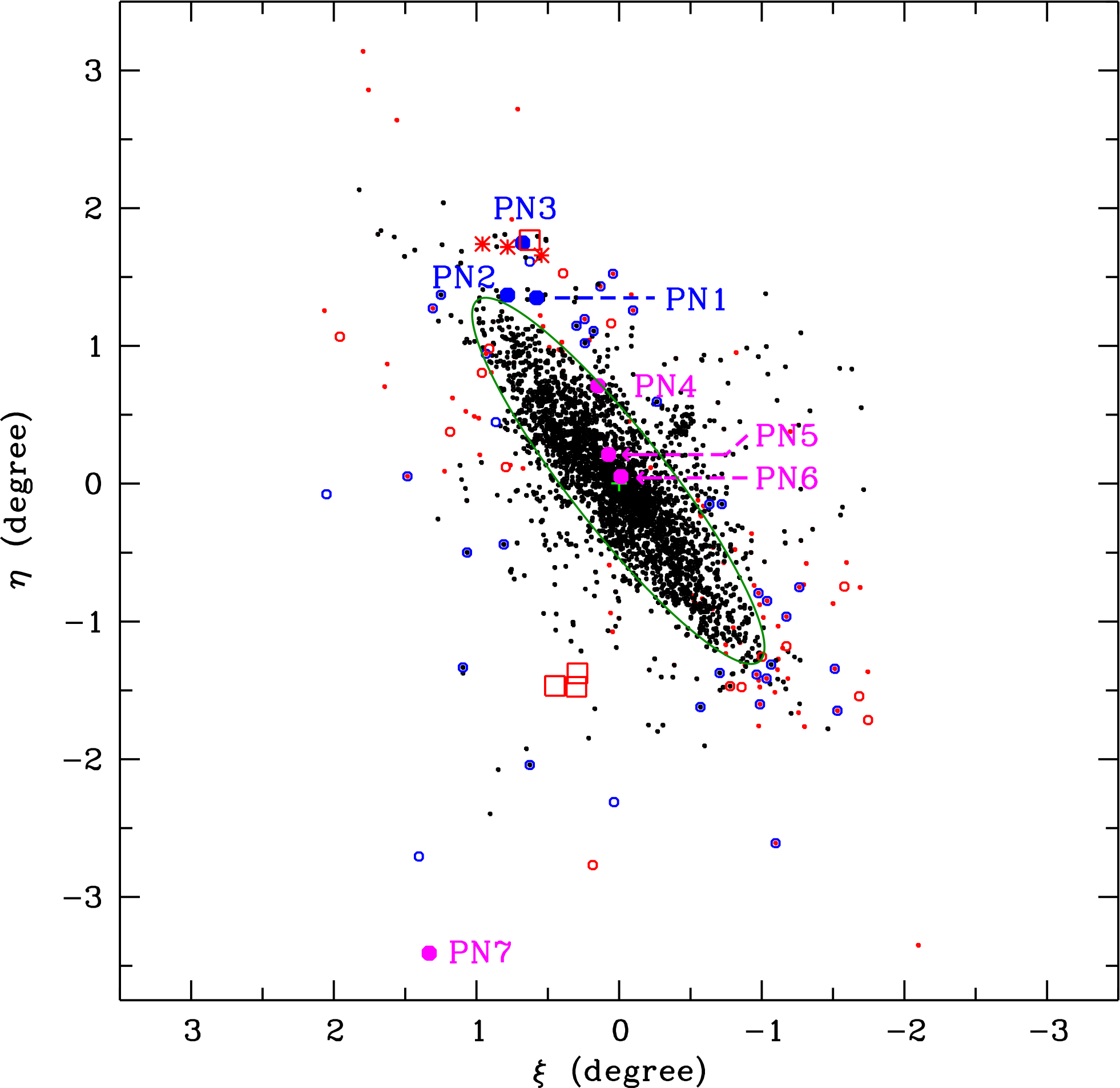}
\caption{
M31 PNe discovered by \citet[][black dots; including those searched by 
\citealt{hkpn04}]{mer06}, \citet[][red dots]{kniazev14} and LAMOST 
(blue open circles, \citealt{yuan10}; red open circles, unpublished). 
The three Northern Spur PNe observed in Paper~I are indicated by red 
asterisks. 
The seven PNe in this study, three in the Northern Spur (blue filled 
circles) and four associated to the extension of the Giant Steam (magenta 
filled circles), are named PN1 to 7.  Red open squares are the \emph{Hubble 
Space Telescope} Advanced Camera for Surveys fields studied by 
\citet{ber15}, one at the Northern Spur and three at the Giant Stream. 
The coordinates $\xi$ and $\eta$ represent the 
M31-based reference frame defined by \citet{hbk91}.  The green ellipse 
is a disk with a 2$^{\circ}$ radius (27.4~kpc) around the center of M31, 
assuming an inclination angle of 77\fdg7 and a position angle of 37\fdg7 
\citep{dev58} for the M31 disk. }
\label{m31_pne}
\end{center}
\end{figure}

\section{Observations and Data Reduction} \label{section2}

\subsection{Target Selection} \label{section2:a}

We selected seven PNe located in or associated with the Northern Spur 
and the extension of the Giant Stream.  Of the seven PNe in our sample, 
six were selected from the catalog of \citet{mer06}, and the other one 
was identified by the 
Large Sky Area Multi-Object Fiber Spectroscopic Telescope (LAMOST; 
\citealt{su98}) during its early commissioning phase \citep{yuan10}. 
Three PNe of our 
sample have been identified as being in the Northern Spur by 
\citet{mer06}, and the other four are kinematically associated with the 
Giant Stream.  Their projected galactocentric distances range from 
0.7--50.2~kpc.  The 
coordinates (right ascension -- R.A., and declination -- Dec.), apparent 
magnitudes $m$($\lambda$5007), radial velocities, and sky-projected 
galactocentric distances (in kpc) are presented in Table~\ref{observ_log}. 
The six PNe selected from \citet{mer06} are relatively bright, with 
$m$($\lambda$5007) $\sim$20.4--21.3.  
If the bright mercury line (Hg~{\sc i} $\lambda$4358) exists, the 
heliocentric velocity of M31 ($\sim$ $-$300~km\,s$^{-1}$) could make the 
[O~{\sc iii}] $\lambda$4363 auroral line in a PN spectrum difficult to 
resolve (e.g., \citealt{kwi12}; Paper~I), and thus a loss of important 
temperature diagnostic for this target.  Given the excellent observing 
condition at ORM in La Palma, where the mercury line is faint, target 
selection were not confined by radial velocities but mainly based on the 
brightness.  The PN newly discovered by LAMOST, which is located 
$\sim$3\fdg65 from the center of M31 (Figure~\ref{m31_pne}), is of 
particular interest because it is both spatially and kinematically 
related to the Giant Stream, and is also the first PN discovered in 
the outer extension of this stream.  These targets are hereafter 
referred to PN1--7 (Table~\ref{observ_log}).

\begin{table*}
\begin{center}
\begin{minipage}{120mm}
\caption{Properties and Observing Log of the Seven PNe}
\label{observ_log}
\begin{tabular}{lcccrrc}
\hline
\hline
PN ID$^{a}$ & 
R.A. & 
Decl. & 
$m$($\lambda$5007)$^{b}$ & 
$v_{\rm helio}$ & 
$R_{\rm gal}\,\,^{c}$ & 
GTC Exp. \\
 & 
(J2000.0) & 
(J2000.0) & 
 & 
(km\,s$^{-1}$) &
(kpc) &
(s) \\
\hline
PN1 (M2445)   & 0:45:52.8 & 42:36:52.9 & 20.65 & $-$149.1 & 20.1 & 4$\times$1250\\
PN2 (M2451)   & 0:46:59.3 & 42:37:58.2 & 20.81 &  $-$80.8 & 21.6 & 4$\times$1250\\
PN3 (M2427)   & 0:46:26.5 & 43:00:43.0 & 20.48 & $-$420.3 & 25.7 & 4$\times$1250\\
PN4 (M77)     & 0:43:32.3 & 41:58:42.5 & 20.99 & $-$579.0 & 9.93 & 4$\times$1200\\
PN5 (M586)    & 0:43:07.9 & 41:28:45.5 & 20.82 & $-$713.5 & 3.05 & 4$\times$1200\\
PN6 (M2750)   & 0:42:40.3 & 41:19:07.4 & 21.23 & $-$766.5 & 0.70 & 4$\times$1200\\
PN7 (LAMOST)  & 0:49:29.0 & 37:50:45.4 & 21.51 & $-$203.0 & 50.2 & 4$\times$2400\\
              &           &            &       &          &      & 1$\times$1200\\
\tableline
\end{tabular}
\begin{description}
\item[$^{a}$] Number in the bracket that follows the PN ID is ID number 
from \citet{mer06} except PN7, which was newly discovered by LAMOST.

\item[$^{b}$] $m$($\lambda$5007) = $-$2.5$\log{F(\lambda5007)}$ $-$ 13.74.

\item[$^{c}$] Sky-projected galactocentric distance.
\end{description}
\end{minipage}
\end{center}
\end{table*}

\subsection{Observations} \label{section2:b}

Spectroscopy of the seven PNe was carried out with OSIRIS (Optical 
System for Imaging and low-intermediate-Resolution Integrated 
Spectroscopy) on the 10.4\,m GTC at ORM in La Palma, Spain.  The 
observations were obtained on seven nights, from of August 23 to 
August 29, $2014$ in service mode.  The excellent observing conditions 
at ORM allow photometric and clear nights with seeing 0\farcs5--1\farcs0. 
The 1\arcsec\-wide long slit and the grism R1000B were used.  The R1000B 
is a 1000 lines~mm$^{-1}$ grism centered at 5455\,{\AA}.  The detector of 
OSIRIS consists of a mosaic of two Marconi CCDs (named CCD1 and CCD2) 
with 2048$\times$4096 pixels each.  The pixel physical size is 
15\,$\mu$m, which corresponds to a scale of 0\farcs127 on the sky.  The 
OSIRIS standard observing mode uses 0\farcs254 binned pixels.  This 
instrument setup enabled us to perform spectroscopy over the wavelength 
range $\sim$3630--7760\,{\AA} with a spectral resolution of 
$\sim$6\,{\AA} (full width at half-maximum, FWHM) at 
2.06\,{\AA}~pixel$^{-1}$.

\begin{figure}
\begin{center}
\includegraphics[width=1.0\columnwidth,angle=0]{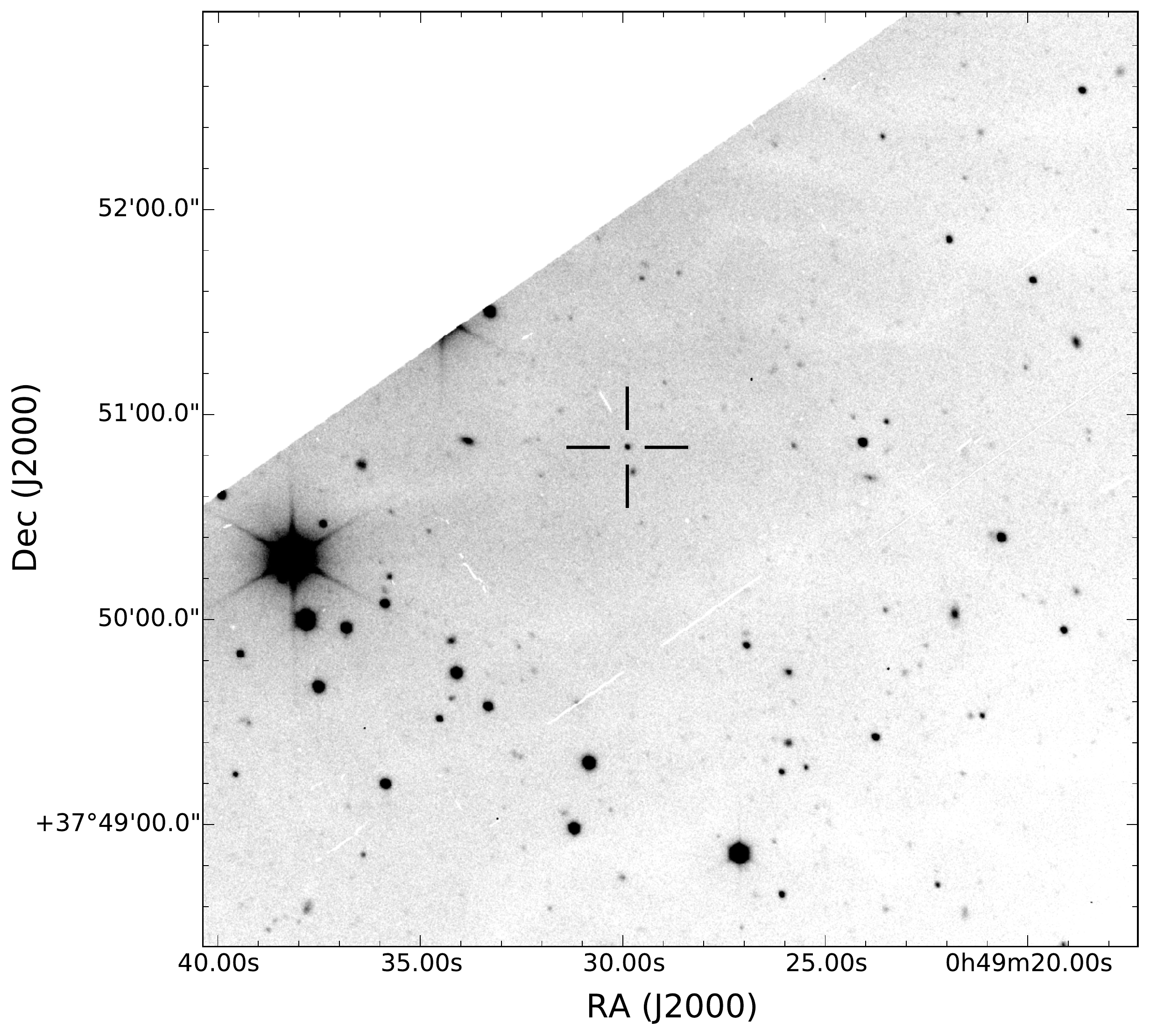}
\caption{Negative greyscale GTC OSIRIS acquisition image of the faint target 
PN7 (marked with the crosshair).  This $g$-band image was taken under 
photometric condition with seeing $\sim$1\arcsec\ at an exposure of 30\,s. }
\label{pn7_acq}
\end{center}
\end{figure}

Except PN6, which is close to the center of M31 and suffers from the 
bright background emission of the galaxy, direct acquisition imaging of 
all PNe was made with an exposure of 20--30s.  Figure~\ref{pn7_acq} is 
the GTC OSIRIS CCD2 acquisition image of PN7, the faintest target in 
our sample (Table~\ref{observ_log}), with an exposure of 30s. 
The GTC exposures are 
summarized in Table~\ref{observ_log}.  In order to avoid light losses 
due to atmospheric diffraction, the long slit was placed along the 
parallactic angle during exposures.  During the observations 
of each night, exposures of the spectrophotometric standard stars 
G191-B2B, Ross640 and G158-100 \citep{oke74,oke90} were made. 
Arc lines of the neon and HgAr lamps were obtained for wavelength 
calibration.  Exposures of the spectrophotometric standards and the 
arc lamps were made right after observations of each PN target.

\subsection{Data Reduction} \label{section2:c}

All data were reduced with standard procedures for long-slit spectra using 
{\sc iraf}\footnote{{\sc iraf}, the Image Reduction and Analysis Facility, 
is distributed by the National Optical Astronomy Observatory, which is 
operated by the Association of Universities for Research in Astronomy under 
cooperative agreement with the National Science Foundation.}.  The raw 
two-dimensional (2D) spectra were first combined, and then bias-subtracted, 
flat-fielded, cosmic-ray removed, and wavelength calibrated using exposures 
of the HgAr lamp.  Geometry rectification, which was used to correct for 
geometry distortion along the slit, was also made during wavelength 
calibration.  The sky background was then fitted along the slit direction 
using spline functions, and subtracted from the 2D spectra.  
The 1D spectrum of each target PN was extracted from the background 
subtracted 2D spectrum.  The 1D spectra were then corrected for 
atmospheric extinction, and flux calibrated using observations of the 
standard stars. 
As an example, Figure~\ref{pn3pn7} shows the fully calibrated 1D 
spectra of PN3 and PN7, the brightest and faintest targets in our 
sample, respectively.

\begin{figure*}[ht]
\begin{center}
\includegraphics[width=2.05\columnwidth,angle=0]{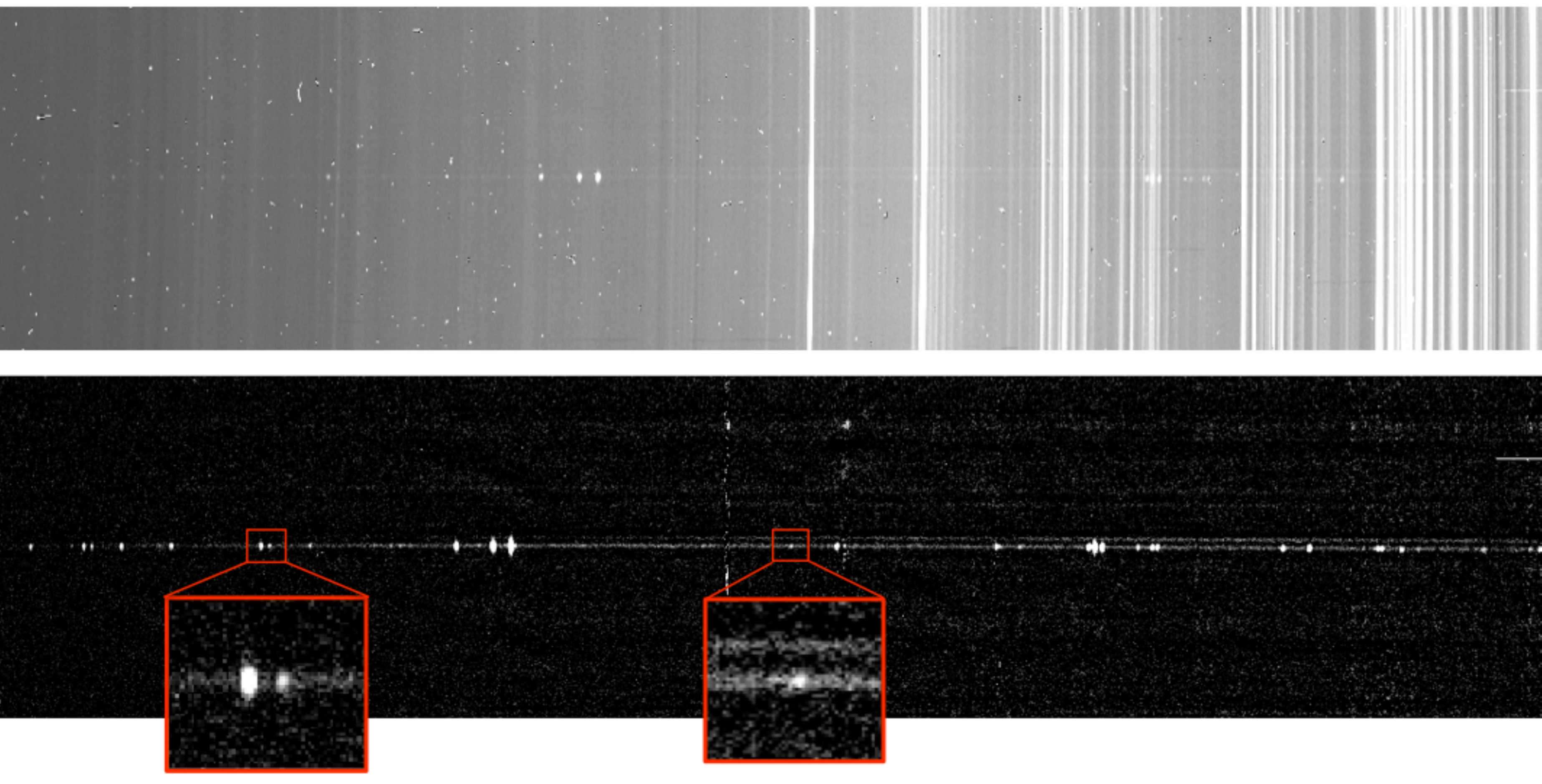}
\caption{GTC OSIRIS 2D spectrum of PN7, the faintest target in our sample. 
The top image is the raw data and the bottom image shows the spectrum 
after cosmic-ray removal, wavelength calibration, geometry rectification, 
and background subtraction.  Images have been trimmed to a size of 
450$\times$2000 pixels.  Please note the shift of emission line positions 
(along the dispersion direction) on CCD after geometry rectification and 
wavelength calibration.  Wavelength coverage is 3630--7750\,{\AA}.  The 
two images, both presented in logarithm, are scaled to different levels. 
The insets of the bottom image show the H$\gamma$ and [O~{\sc iii}] 
$\lambda$4363 lines (left) and the [N~{\sc ii}] $\lambda$5755 line 
(right). }
\label{2d_spec}
\end{center}
\end{figure*}

Much effort was made to subtract the sky background.  The excellent 
observing condition at ORM enables detection of the [O~{\sc iii}] 
$\lambda$4363 auroral line even in the faintest target PN of our sample 
(Figure~\ref{2d_spec}, top panel).  However, the sky background in the red 
part ($>$6000\,{\AA}) is relatively strong.  Besides, intensity of sky 
lines is inhomogeneous along the long slit, and this inhomogeneity becomes 
significant for the strong sky lines such as [O~{\sc i}] 
$\lambda\lambda$5577,\,6300,\,6363.  We used multiple cubic-spline 
functions to carefully fit the background emission along the slit 
direction, and then subtract it from the wavelength-calibrated (also 
geometry-rectified) 2D spectrum.  Subtraction of the background emission 
is generally satisfactory (e.g., Figure~\ref{2d_spec}, bottom panel) 
except PN6, which is close to the center of M31 and thus its spectrum 
is affected by the strong galactic background emission.  Despite much 
effort, the signal-to-noise (S/N) ratio of the 1D spectrum of PN6 is 
lower than those of the other six targets.

The [O~{\sc iii}] $\lambda$4363 auroral line is seen in the reduced 2D 
spectra of all seven PNe (e.g., Figure~\ref{2d_spec}, bottom panel). 
The [N~{\sc ii}] $\lambda$5755 auroral line is also detected in the 
spectra of relatively bright PNe.  We also managed to discern the 
nebular emission of the [O~{\sc i}] $\lambda\lambda$6300,\,6363 lines 
in the spectra of all targets except PN6.  The [S~{\sc iii}] 
$\lambda$6312 auroral line, which is close to [O~{\sc i}] $\lambda$6300, 
was detected in PN1, PN2, PN3 and PN7 (Figure~\ref{pn3pn7}).  The former 
three PNe are bright and located in the outskirt Northern Spur, and the 
latter one in the outer stream of M31, far ($\sim$3\fdg65; see 
Figure~\ref{m31_pne}) from the galactic center.  These four PNe also 
suffer much less from the galactic background emission than the other 
three.

We noticed the presence of second-order contamination in the final 1D 
spectra of the targets (Figure~\ref{pn3pn7}), although the R1000B 
grism was claimed to be free of such effect.  Our derived response 
curve shows that the second-order contamination is obvious above 
6300\,{\AA}.  We corrected the red part (6200-7760\,{\AA}) of the 
response curve by carrying out global fitting using a five-order 
polynomial.  The efficiency curve of R1000B given by the OSIRIS User 
Manual \footnote{URL 
http://www.gtc.iac.es/instruments/osiris/media/OSIRIS-USER-MANUAL$\_$v3$\_$1.pdf} 
was used as a template.  The final corrected R1000B response curve is 
safe for flux calibration (A. Cabrera-Lavers, GTC, private communications).

\subsection{Extinction Correction} \label{section2:d}

The logarithmic extinction parameter at H$\beta$, $c$(H$\beta$), was derived 
for the seven M31 PNe by comparing the observed H~{\sc i} Balmer line ratios 
with the predicted Case~B values.  
It is worth noting that some low-order H~{\sc i} Balmer lines are blended
with He~{\sc ii} lines: H$\gamma$ is blended with He~{\sc ii} $\lambda$4338
(4f\,$^{2}$F$^{\rm o}$ -- 10g\,$^{2}$G), H$\beta$ is blended with He~{\sc ii}
$\lambda$4859 (4f\,$^{2}$F$^{\rm o}$ -- 8g\,$^{2}$G), and H$\alpha$ with
He~{\sc ii} $\lambda$6560 (4f\,$^{2}$F$^{\rm o}$ -- 6g\,$^{2}$G).  Flux
contributions of these He~{\sc ii} lines are negligible ($<$2\%).
The other low-order Balmer lines were not used for extinction correction
because H~{\sc i} $\lambda$3889 ($n$ = 2--8) is blended with the He~{\sc i}
$\lambda$3888 (2s\,$^{3}$S -- 3p\,$^{3}$P$^{\rm o}$) line, H~{\sc i}
$\lambda$3970 ($n$ = 2--7) is blended with [Ne~{\sc iii}] $\lambda$3967
(2p$^{4}$\,$^{3}$P$_{1}$ -- $^{1}$D$_{2}$), and H$\delta$ ($\lambda$4101)
is probably blended with the N~{\sc iii} $\lambda\lambda$4097,\,4103
(3s\,$^{2}$S -- 3p\,$^{2}$P$^{\rm o}$) lines, 
although the flux contribution of the latter is expected to be 
negligible.

The average $c$(H$\beta$) value derived from the H$\alpha$/H$\beta$ and 
H$\gamma$/H$\beta$ ratios, as adopted for all PNe, are presented in 
Table~\ref{lines}.  Here the theoretical H~{\sc i} Balmer line ratios were 
adopted from \citet{sh95}, assuming an electron temperature of 10\,000~K 
and a density of 10$^{4}$~cm$^{-3}$.  Given that the H~{\sc i} recombination
line ratios are insensitive to the electron temperature and essentially 
independent on the density, this assumption of $T_\mathrm{e}$ and 
$N_\mathrm{e}$ is reasonable.  The observed line fluxes were dereddened 
by 

\begin{equation}
\label{deredden}
I(\lambda) = 10^{c(\rm{H}\beta) {\it f}(\lambda)} F(\lambda),
\end{equation}
where $f(\lambda)$ is the extinction curve adopted from \citet{ccm89} 
with a total-to-selective extinction ratio $R_{\rm V}$ = 3.1.  
The effects on the extinction-corrected line intensities caused 
by the use of different reddening curves \citep[e.g.,][]{sm79,ccm89,clay15} 
was assessed.  They resulted in very small differences in the dereddened 
intensities of emission lines in the wavelength range 3630--7760\,{\AA} 
covered by the R1000B grism of GTC OSIRIS. 
The observed fluxes and the extinction-corrected intensities of the 
emission lines detected in all seven M31 PNe, both normalized to 
H$\beta$ = 100, are presented in Table~\ref{lines}.  The total H$\beta$ 
fluxes of the targets, as integrated in the extracted spectra, are also 
presented.  Given that the slit width (1\farcs0) we used is larger than 
the typical seeing (0\farcs6--0\farcs8) and that the slit is always 
placed along the parallactic angle during exposures, no significant 
light loss in H$\beta$ is expected.

\section{Results} \label{section3}

\subsection{Relative Line Intensities} \label{section3:a}

We present in Table~\ref{lines} the normalized (to H$\beta$ = 100) 
intensities and measurement errors of the emission lines. 
For brighter PNe, the typical uncertainties in fluxes of the [O~{\sc 
iii}] $\lambda$4363 auroral line is less than 10\%.  For the PN1, PN2 
and PN3, the relatively bright targets in our sample, the S/N ratios of 
the $\lambda$4363 line are $\sim$30, 21 and $>$50, respectively.  For the 
faintest target PN7, S/N([O~{\sc iii}]~$\lambda$4363) $\sim$10.  Electron
temperature diagnostics using the [O~{\sc iii}] ($\lambda$4959 + 
$\lambda$5007)/$\lambda4363$ line ratio are presented in 
Section~\ref{section3:c}.

Some emission lines detected in our spectra have blending issues.  We 
corrected the flux of the [Ne~{\sc iii}] $\lambda$3967 line for the 
blended H~{\sc i} $\lambda$3970 line, using the theoretical hydrogen 
line ratio of \citet{sh95}. 
However, for all targets, the corrected [Ne~{\sc iii}] $\lambda$3967 
(dereddened) line flux still yields higher Ne$^{2+}$/H$^{+}$ abundance 
ratios than that derived from [Ne~{\sc iii}] $\lambda$3868, which is 
free of line blending. 
The [Ar~{\sc iv}] $\lambda$4711 line is blended with He~{\sc i} 
$\lambda$4713 (2p\,$^{3}$P$^{\rm o}$ -- 4s\,$^{3}$S), which contributes 
30--40\% to the total flux of the [Ar~{\sc iv}] line for all seven 
targets. 
The He~{\sc i} line flux was estimated using the theoretical 
Case~B He~{\sc i} $\lambda$4713/$\lambda$4471 ratio calculated by 
\citet{porter12}, whose calculation for this ratio only differs from 
that of \citet{bss99} by less than 0.5\%. 
The [S~{\sc iii}] $\lambda$6312 line is blended with He~{\sc ii} 
$\lambda$6311 (5g\,$^{2}$G -- 16h\,$^{2}$H$^{\rm o}$), which typically 
contributes less than 1\% to the total flux.  For PN3, whose He~{\sc 
ii} $\lambda$4686 is the strongest among the targets, flux contribution 
of the He~{\sc ii} $\lambda$6311 line is close to 5\%.

\begin{figure*}
\begin{center}
\includegraphics[width=17.5cm,angle=0]{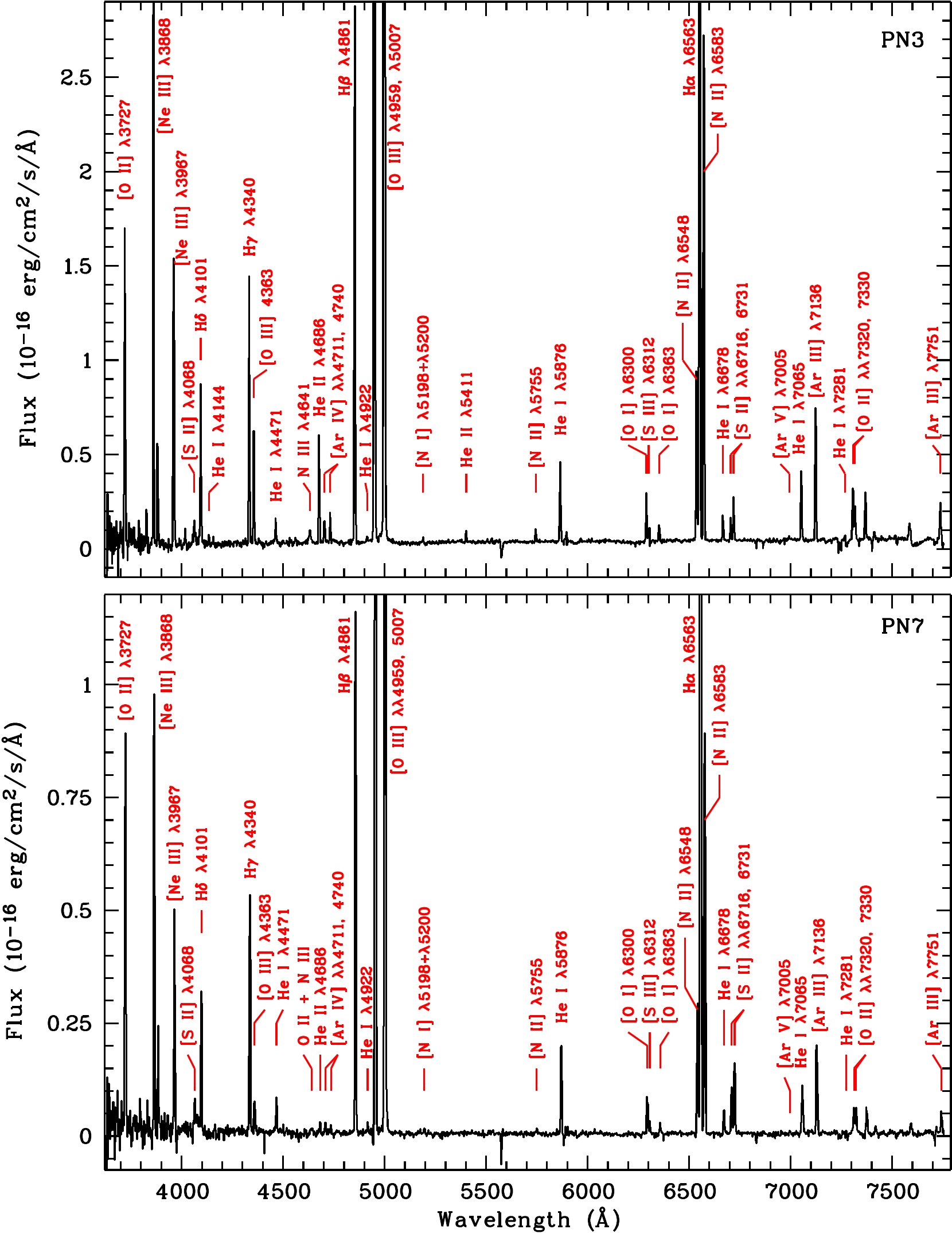}
\caption{
GTC OSIRIS 1D spectrum of PN3 (top) and PN7 (bottom), the brightest 
and the faintest targets in our sample (Table~\ref{observ_log}), 
respectively.  All important emission lines are labeled.  Extinction 
has not been corrected for.  The features between the [O~{\sc ii}] 
$\lambda\lambda$7320,7330 and [Ar~{\sc iii}] $\lambda$7751 lines are 
the second-order contamination, probably belong to [Ne~{\sc iii}] 
$\lambda$3868, H~{\sc i} $\lambda$3889 and [Ne~{\sc iii}] 
$\lambda$3967. }
\label{pn3pn7}
\end{center}
\end{figure*}

Some permitted lines of heavy elements were detected (Table~\ref{lines}). 
These lines are faint, with a typical intensity $<$5\% of H$\beta$.  The 
N~{\sc iii} $\lambda$4641 (M2 3p\,$^{2}$P$^{\rm o}_{3/2}$ -- 
3d\,$^{2}$D$_{5/2}$) line was well detected in the spectrum 
of PN3 (Figure~\ref{pn3pn7}).  This line is blended with N~{\sc iii} 
$\lambda$4634,\,4642 of the same multiplet, and is probably blended with 
the O~{\sc ii} M1 3s\,$^{4}$P -- 3p\,$^{4}$D$^{\rm o}$ lines (see for 
example, Figure~$1$ in \citealt{liu95} or Figure~$32$ in \citealt{fl13}). 
The dominant excitation mechanism of the [N~{\sc iii}] $\lambda$4641 line 
is the starlight and/or Bowen fluorescence (e.g., \citealt{bow35}; 
\citealt{gra76}).  Thus this line is unsuitable for the ionic abundance 
(N$^{3+}$/H$^{+}$) calculation.  Carbon emission lines were detected in 
the spectrum of PN2, as discussed in Section~\ref{section3:b} below.

The uncertainties of the emission line intensities were derived by 
adding quadratically the uncertainties associated with the Poisson 
noise of the line and the CCD and background noise.  The former scales 
with the electron counts of the line, and the latter scales with the 
local noise and the full line width.  The error contributions from 
systematic effects, such as flux calibration, subtraction of the 
background level, determination of reddening, or line blending, are 
expected to be much smaller and were not taken into account.  Our flux 
calibration is reliable as the extinction parameter $c$(H$\beta$) 
derived from the H$\gamma$/H$\beta$ ratio generally agrees with that 
derived from H$\alpha$/H$\beta$.  The important nebular emission lines 
of our PNe are generally not much affected by the sky lines (e.g., the 
mercury line Hg~{\sc i} $\lambda$4358 is faint at ORM, La Palma, and 
the detection of the [O~{\sc iii}] $\lambda$4363 auroral is not much 
affected).  Moreover, the background subtraction in the 2D spectra 
seemed flawless (e.g., Figure~\ref{2d_spec}; see also the description 
of data reduction in Section~\ref{section2:c}).  The errors introduced 
by reddening are also small, less than 5\% as discussed in 
Section~\ref{section2:d}.  Finally, line blending was also considered 
when analyzing the emission lines.  Fortunately, errors due to this 
effect are mostly negligible for the lines that are critical for plasma 
diagnostics and abundance determinations (Table~\ref{lines}), except 
for the [Ar~{\sc iv}] $\lambda$4711 line.

\subsection{Carbon Emission Lines} \label{section3:b}

\begin{figure*}
\begin{center}
\includegraphics[width=7.2cm,angle=-90]{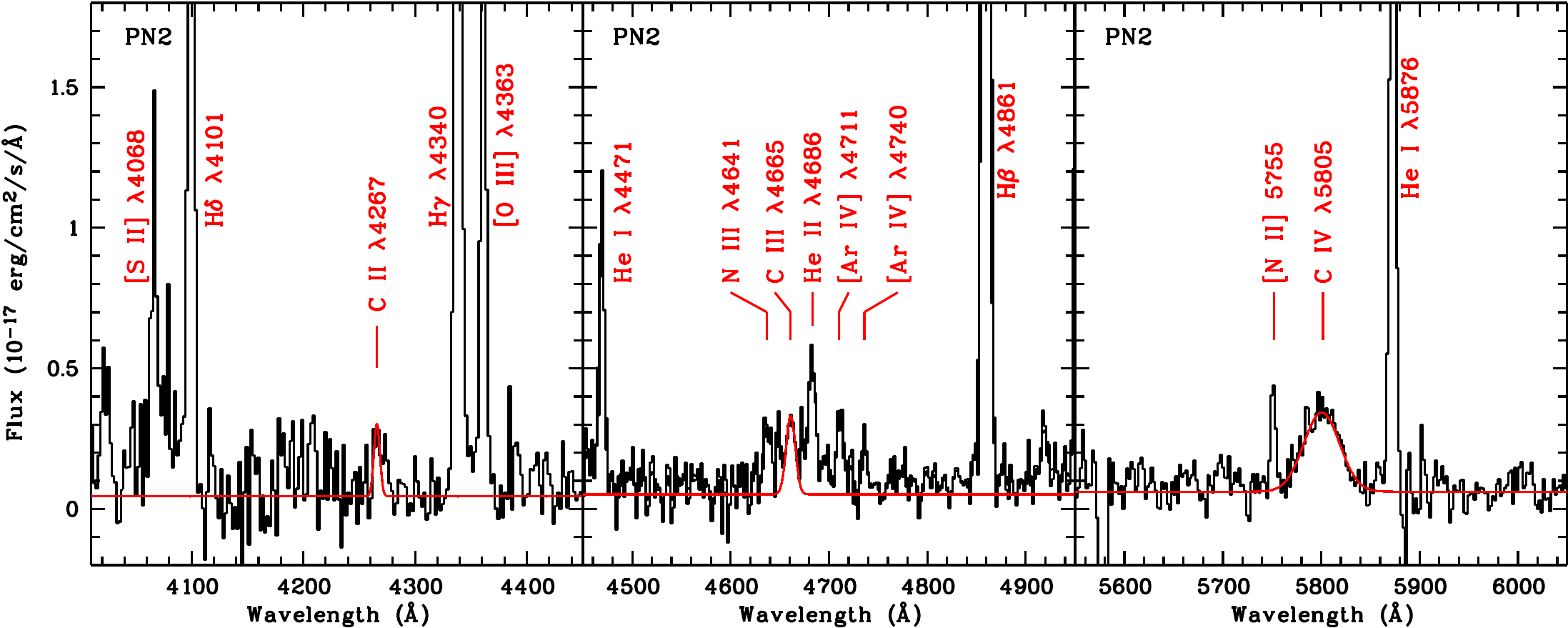}
\caption{The carbon lines detected in the spectrum of PN2:  C~{\sc ii}
$\lambda$4267 (left), C~{\sc iii} $\lambda$4665 (middle), and C~{\sc iv}
$\lambda$5805 (right).  The broad C~{\sc iv} $\lambda$5805 feature is
emitted by a Wolf-Rayet type central star.  The red continuous curves are
Gaussian profile fits to the carbon lines, with FWHM=43\,{\AA} for the
$\lambda$5805 line.  An FWHM of 6\,{\AA} is assumed for the $\lambda$4267
line and $\sim$11\,{\AA} for $\lambda$4665. }
\label{pn2_carbon}
\end{center}
\end{figure*}

Very broad C~{\sc iv} $\lambda$5805 (M1 3s\,$^{2}$S -- 3p\,$^{2}$P$^{\rm 
o}$) line, with FWHM $\sim$43\,{\AA}, was detected in the spectrum of PN2 
(Figure~\ref{pn2_carbon}, right).  This line is a 
blend of two fine-structure components 
$\lambda\lambda$5801.33,\,5811.98\footnote{The wavelengths of these two 
C~{\sc iv} lines are 
adopted from the laboratory wavelengths in the atomic spectral line lists 
compiled by \citet{hh95}}, and can be attributed to the emission of a 
Wolf-Rayet (WR) type central star.  The same feature has been detected in 
the spectra of another two PNe in the outer disk of M31 by 
\citet[][Figure~2]{balick13}.  The extinction-corrected flux of 
the C~{\sc iv} $\lambda$5805 line in our PN2 is 
1.20$\times$10$^{-16}$~erg\,cm$^{-2}$\,s$^{-1}$, with an uncertainty of 
$\sim$10\%.  This flux is generally in line with those observed by 
\citet{balick13}.

Also detected in the spectrum of PN2 are the C~{\sc iii} $\lambda$4665 
(M5 3s$^{\prime}$\,$^{3}$P$^{\rm o}_{2}$ -- 3p$^{\prime}$\,$^{3}$P$_{2}$) 
and C~{\sc ii} $\lambda$4267 (M6 3d\,$^{2}$D -- 4f\,$^{2}$F$^{\rm o}$) 
lines.  The high-excitation C~{\sc iii} $\lambda$4665 line might be due to 
emission of the WR central star.  A Gaussian profile with FWHM 
$\sim$11\,{\AA} seems to well fit this line (Figure~\ref{pn2_carbon}, 
middle), but its accurate profile is difficult to discern due to noise and 
line blending.  C~{\sc iii} $\lambda$4665 is probably blended with the 
C~{\sc iv} $\lambda$4658 (M8 5g\,$^{2}$G -- 6h\,$^{2}$H$^{\rm o}$) and 
C~{\sc iii} $\lambda$4649 (M1 3s\,$^{3}$S -- 3p\,$^{3}$P$^{\rm o}$) 
recombination lines.

The C~{\sc ii} $\lambda$4267 optical recombination line (ORL) is produced 
by radiative recombination only, and thus is due to nebular emission. 
Its line width, although difficult to discern due to noise and possible 
blending (Figure~\ref{pn2_carbon}, left), seems narrower than that of 
the C~{\sc iv} $\lambda$5805 line.  The extinction-corrected flux of 
the $\lambda$4267 line is 
$\sim$2.1$\times$10$^{-17}$~erg\,cm$^{-2}$\,s$^{-1}$, with an uncertainty 
over 50\%.  This flux value is close to that observed in one of the two 
PNe of \citet{balick13}, which is $\sim$1.8$\times$10$^{-17}$
erg\,cm$^{-2}$\,s$^{-1}$.  The C~{\sc ii} $\lambda$4267 line was used to 
calculate the C$^{2+}$/H$^{+}$ abundance ratio in Section~\ref{section3:d}.

\begin{table*}
\begin{center}
\caption{Fluxes and Intensities$^{a}$}
\label{lines}
\begin{tabular}{lllcccccccc}
\hline\hline
\rule[-1mm]{0mm}{5.0mm}\\
Ion  & $\lambda$  & Transition & 
\multicolumn{2}{c}{\underline{~~~~~~~PN1~~~~~~~}} & 
\multicolumn{2}{c}{\underline{~~~~~~~PN2~~~~~~~}} & 
\multicolumn{2}{c}{\underline{~~~~~~~PN3~~~~~~~}} & 
\multicolumn{2}{c}{\underline{~~~~~~~PN4~~~~~~~}} \\ 
     &  (\AA)     &            & 
$F$($\lambda$) & $I$($\lambda$) & 
$F$($\lambda$) & $I$($\lambda$) & 
$F$($\lambda$) & $I$($\lambda$) & 
$F$($\lambda$) & $I$($\lambda$) \\
\hline
\rule[-1mm]{0mm}{5.0mm}\\
$[$O~{\sc ii}$]$   &  3727$^{b}$ & 2p$^{3}$\,$^{4}$S$^{\rm o}$ -- 2p$^{3}$\,$^{2}$D$^{\rm o}$              &   47.2   &  54.6$\pm$5.4   &   79.3   &  91.8$\pm$9.1   &   58.2   &  66.4$\pm$6.6   &   56.2   &  67.6$\pm$6.7 \\
H~{\sc i}          &  3798       & 2p\,$^{2}$P$^{\rm o}$ -- 10d\,$^{2}$D                                   &    4.28  &  4.92$\pm$1.35  &    8.42  &  9.69$\pm$2.60  &    3.01  &  3.42$\pm$0.94  &    4.51  &  5.38$\pm$1.48\\
H~{\sc i}          &  3835       & 2p\,$^{2}$P$^{\rm o}$ -- 9d\,$^{2}$D                                    &    5.84  &  6.68$\pm$1.70  &    6.06  &  6.94$\pm$1.74  &    4.88  &  5.52$\pm$1.38  &    3.22  &  3.82$\pm$1.00\\
$[$Ne~{\sc iii}$]$ &  3868       & 2p$^{4}$\,$^{3}$P$_{2}$ -- 2p$^{4}$\,$^{1}$D$_{2}$                      &   74.4   &  84.8$\pm$4.2   &  112     &   128$\pm$6     &  104     &   117$\pm$6     &   69.1   &  81.7$\pm$4.0 \\
H~{\sc i}          &  3889$^{c}$ & 2p\,$^{2}$P$^{\rm o}$ -- 8d\,$^{2}$D                                    &   16.2   &  18.4$\pm$2.0   &   16.0   &  18.2$\pm$2.0   &   16.8   &  19.0$\pm$2.0   &   14.5   &  17.1$\pm$1.9 \\
$[$Ne~{\sc iii}$]$ &  3967$^{d}$ & 2p$^{4}$\,$^{3}$P$_{1}$ -- 2p$^{4}$\,$^{1}$D$_{2}$                      &   41.2   &  46.5$\pm$3.6   &   52.3   &  59.1$\pm$4.6   &   53.8   &  60.0$\pm$4.6   &   38.3   &  44.6$\pm$3.5 \\
He~{\sc i}         &  4026       & 2p\,$^{3}$P$^{\rm o}$ -- 5d\,$^{3}$D                                    &    2.21  &  2.47$\pm$0.50  &    3.57  &  4.00$\pm$0.83  &    1.89  &  2.10$\pm$0.42  &    2.57  &  2.97$\pm$0.60\\
$[$S~{\sc ii}$]$   &  4068$^{e}$ & 3p$^{3}$\,$^{4}$S$^{\rm o}_{3/2}$ -- 3p$^{3}$\,$^{2}$P$^{\rm o}_{3/2}$  &    4.56  &  5.10$\pm$1.78  &    7.78  &  8.67$\pm$2.20  &    5.99  &  6.60$\pm$1.60  &    3.13  &  3.60$\pm$1.10\\
H~{\sc i}          &  4101       & 2p\,$^{2}$P$^{\rm o}$ -- 6d\,$^{2}$D                                    &   26.8   &  30.0$\pm$3.3   &   29.1   &  32.4$\pm$3.6   &   29.5   &  32.4$\pm$3.6   &   28.9   &  33.0$\pm$3.6 \\
He~{\sc i}         &  4144       & 2p\,$^{1}$P$^{\rm o}_{1}$ -- 6d\,$^{1}$D$_{2}$                          &    2.24  &  2.50$\pm$0.74  &    ...   &  ...            &    1.15  &  1.26$\pm$0.38  &    ...   &  ...          \\
C~{\sc ii}         &  4267       & 3d\,$^{2}$D -- 4f\,$^{2}$F$^{\rm o}$                                    &    ...   &  ...            &    1.60  &  1.74$\pm$0.43  &    ...   &  ...            &    ...   &  ...          \\
H~{\sc i}          &  4340$^{f}$ & 2p\,$^{2}$P$^{\rm o}$ -- 5d\,$^{2}$D                                    &   44.4   &  47.7$\pm$2.1   &   45.7   &  49.1$\pm$2.2   &   43.6   &  46.5$\pm$2.0   &   42.3   &  46.3$\pm$2.0 \\
$[$O~{\sc iii}$]$  &  4363       & 2p$^{2}$\,$^{1}$D$_{2}$ -- 2p$^{2}$\,$^{1}$S$_{0}$                      &    8.28  &   8.9$\pm$1.0   &   11.9   &  12.8$\pm$1.5   &   18.1   &  19.2$\pm$2.2   &    8.63  &  9.42$\pm$1.13\\
He~{\sc i}         &  4471       & 2p\,$^{3}$P$^{\rm o}$ -- 4d\,$^{3}$D                                    &    5.58  &  5.88$\pm$0.94  &    6.18  &  6.52$\pm$1.04  &    5.10  &  5.35$\pm$0.85  &    6.11  &  6.53$\pm$1.05\\
N~{\sc iii}        &  4641$^{g}$ & 3p\,$^{2}$P$^{\rm o}_{3/2}$ -- 3d\,$^{2}$D$_{5/2}$                      &    0.83  &   0.9$\pm$0.5   &    2.03  &  2.09$\pm$0.52  &    3.57  &  3.67$\pm$0.90  &    ...   &  ...          \\
C~{\sc iii}        &  4665$^{h}$ & 3s$^{\prime}$\,$^{3}$P$^{\rm o}_{2}$ -- 3p$^{\prime}$\,$^{3}$P$_{2}$    &    ...   &  ...            &    2.77  &  2.84$\pm$0.71  &    ...   &  ...            &    ...   &  ...          \\
He~{\sc ii}        &  4686       & 3d\,$^{2}$D -- 4f\,$^{2}$F$^{\rm o}$                                    &    0.74  &  0.76$\pm$0.30  &    4.00  &  4.10$\pm$0.60  &   18.6   &  19.0$\pm$1.7   &    1.36  &  1.40$\pm$0.43\\
$[$Ar~{\sc iv}$]$  &  4711$^{i}$ & 3p$^{3}$\,$^{4}$S$^{\rm o}_{3/2}$ -- 3p$^{3}$\,$^{2}$D$^{\rm o}_{5/2}$  &    1.13  &  1.15$\pm$0.35  &    1.21  &  1.24$\pm$0.20  &    4.41  &  4.49$\pm$0.67  &    2.18  &  2.24$\pm$0.68\\
$[$Ar~{\sc iv}$]$  &  4740       & 3p$^{3}$\,$^{4}$S$^{\rm o}_{3/2}$ -- 3p$^{3}$\,$^{2}$D$^{\rm o}_{3/2}$  &    1.03  &  1.04$\pm$0.34  &    1.01  &  1.02$\pm$0.20  &    5.19  &  5.27$\pm$0.80  &    1.97  &  2.01$\pm$0.60\\
H~{\sc i}          &  4861       & 2p\,$^{2}$P$^{\rm o}$ -- 4d\,$^{2}$D                                    &  100     &   100           &  100     &   100           &  100     &   100           &  100     &   100         \\
He~{\sc i}         &  4922       & 2p\,$^{1}$P$^{\rm o}_{1}$ -- 4d\,$^{1}$D$_{2}$                          &    2.0   &   2.0$\pm$0.8   &    1.97  &  1.96$\pm$0.77  &    1.20  &  1.20$\pm$0.47  &    1.90  &  1.89$\pm$0.75\\
$[$O~{\sc iii}$]$  &  4959       & 2p$^{2}$\,$^{1}$P$_{1}$ -- 2p$^{2}$\,$^{1}$D$_{2}$                      &  345     &   342$\pm$17    &  434     &   429$\pm$21    &  482     &   478$\pm$23    &  321     &   317$\pm$15  \\
$[$O~{\sc iii}$]$  &  5007       & 2p$^{2}$\,$^{1}$P$_{2}$ -- 2p$^{2}$\,$^{1}$D$_{2}$                      & 1042     &  1025$\pm$31    & 1297     &  1275$\pm$38    & 1414     &  1393$\pm$42    &  956     &   937$\pm$28  \\
$[$N~{\sc i}$]$    &  5198$^{j}$ & 2p$^{3}$\,$^{4}$S$^{\rm o}_{3/2}$ -- 2p$^{3}$\,$^{2}$D$^{\rm o}_{3/2}$  &    ...   &  ...            &    1.80  &  1.73$\pm$0.50  &    0.67  &  0.65$\pm$0.21  &    ...   &  ...          \\
He~{\sc ii}        &  5411       & 4f\,$^{2}$F$^{\rm o}$ -- 7g\,$^{2}$G                                    &    ...   &  ...            &    ...   &  ...            &    1.81  &  1.72$\pm$0.34  &    ...   &  ...          \\
$[$Cl~{\sc iii}$]$ &  5517       & 3p$^{3}$\,$^{4}$S$^{\rm o}_{3/2}$ -- 3p$^{3}$\,$^{2}$D$^{\rm o}_{5/2}$  &    ...   &  ...            &    ...   &  ...            &    0.48  &  0.45$\pm$0.20  &    ...   &  ...          \\
$[$Cl~{\sc iii}$]$ &  5537       & 3p$^{3}$\,$^{4}$S$^{\rm o}_{3/2}$ -- 3p$^{3}$\,$^{2}$D$^{\rm o}_{3/2}$  &    ...   &  ...            &    ...   &  ...            &    0.72  &  0.68$\pm$0.21  &    ...   &  ...          \\
$[$N~{\sc ii}$]$   &  5755       & 2p$^{2}$\,$^{1}$D$_{2}$ -- 2p$^{2}$\,$^{1}$S$_{0}$                      &    1.90  &  1.76$\pm$0.50  &    2.06  &  1.89$\pm$0.60  &    2.64  &  2.45$\pm$0.50  &    1.53  &  1.38$\pm$0.42\\
C~{\sc iv}         &  5805       & 3s\,$^{2}$S -- 3p\,$^{2}$P$^{\rm o}$                                    &    ...   &  ...            &   11.6   &  10.7$\pm$0.8   &    ...   &  ...            &    ...   &  ...          \\
He~{\sc i}         &  5876       & 2p\,$^{3}$P$^{\rm o}$ -- 3d\,$^{3}$D                                    &   19.1   &  17.4$\pm$1.9   &   20.8   &  18.9$\pm$2.1   &   16.3   &  15.0$\pm$1.6   &   17.5   &  15.6$\pm$1.7 \\
$[$O~{\sc i}$]$    &  6300       & 2p$^{4}$\,$^{3}$P$_{2}$ -- 2p$^{4}$\,$^{1}$D$_{2}$                      &    4.10  &  3.64$\pm$0.94  &   10.8   &  9.60$\pm$2.40  &    8.82  &  7.93$\pm$2.00  &    5.42  &  4.68$\pm$1.20\\
$[$S~{\sc iii}$]$  &  6312$^{k}$ & 3p$^{2}$\,$^{1}$D$_{2}$ -- 3p$^{2}$\,$^{1}$S$_{0}$                      &    1.37  &  1.21$\pm$0.37  &    2.55  &  2.26$\pm$0.69  &    1.41  &  1.26$\pm$0.25  &    ...   &  ...          \\
$[$O~{\sc i}$]$    &  6363       & 2p$^{4}$\,$^{3}$P$_{1}$ -- 2p$^{4}$\,$^{1}$D$_{2}$                      &    1.84  &  1.63$\pm$0.73  &    4.52  &  4.00$\pm$1.79  &    3.32  &  2.97$\pm$1.33  &    2.30  &  1.98$\pm$0.88\\
$[$N~{\sc ii}$]$   &  6548       & 2p$^{2}$\,$^{3}$P$_{1}$ -- 2p$^{2}$\,$^{1}$D$_{2}$                      &   21.0   &  18.4$\pm$2.2   &   38.0   &  33.3$\pm$4.0   &   34.0   &  30.2$\pm$2.3   &   22.6   &  19.1$\pm$2.3 \\
H~{\sc i}          &  6563       & 2p\,$^{2}$P$^{\rm o}$ -- 2d\,$^{2}$D                                    &  334     &   293$\pm$1     &  356     &   310$\pm$1     &  328     &   291$\pm$1     &  341     &   288$\pm$1   \\
$[$N~{\sc ii}$]$   &  6583       & 2p$^{2}$\,$^{3}$P$_{2}$ -- 2p$^{2}$\,$^{1}$D$_{2}$                      &   64.7   &  56.5$\pm$6.4   &  112     &  97.9$\pm$10.7  &   99.6   &  88.2$\pm$5.0   &   68.2   &  57.5$\pm$6.3 \\
He~{\sc i}         &  6678$^{l}$ & 2p\,$^{1}$P$^{\rm o}_{1}$ -- 3d\,$^{1}$D$_{2}$                          &    5.76  &  5.01$\pm$0.75  &    6.88  &  5.97$\pm$0.90  &    4.74  &  4.17$\pm$0.62  &    5.46  &  4.58$\pm$0.68\\
$[$S~{\sc ii}$]$   &  6716       & 3p$^{3}$\,$^{4}$S$^{\rm o}_{3/2}$ -- 3p$^{3}$\,$^{2}$D$^{\rm o}_{5/2}$  &    2.17  &  1.88$\pm$0.20  &    5.21  &  4.51$\pm$0.50  &    4.34  &  3.82$\pm$0.43  &    7.24  &  6.05$\pm$0.67\\
$[$S~{\sc ii}$]$   &  6731       & 3p$^{3}$\,$^{4}$S$^{\rm o}_{3/2}$ -- 3p$^{3}$\,$^{2}$D$^{\rm o}_{3/2}$  &    4.00  &  3.45$\pm$0.42  &    9.24  &  8.00$\pm$1.01  &    8.39  &  7.37$\pm$0.90  &    9.97  &  8.32$\pm$0.92\\
$[$Ar~{\sc v}$]$   &  7005       & 3p$^{2}$\,$^{3}$P$_{2}$ -- 3p$^{2}$\,$^{1}$D$_{2}$                      &    ...   &  ...            &    0.70  &  0.60$\pm$0.12  &    0.77  &  0.67$\pm$0.13  &    ...   &  ...          \\
He~{\sc i}         &  7065       & 2p\,$^{3}$P$^{\rm o}$ -- 3s\,$^{3}$S                                    &   13.9   &  11.8$\pm$1.5   &   15.7   &  13.3$\pm$1.7   &   13.2   &  11.3$\pm$1.4   &    9.24  &  7.52$\pm$1.00\\
$[$Ar~{\sc iii}$]$ &  7136       & 3p$^{4}$\,$^{3}$P$_{2}$ -- 3p$^{4}$\,$^{1}$D$_{2}$                      &   18.4   &  15.5$\pm$2.0   &   19.1   &  16.1$\pm$2.1   &   24.7   &  21.2$\pm$2.7   &   18.8   &  15.2$\pm$2.0 \\
He~{\sc i}         &  7281       & 2p\,$^{1}$P$^{\rm o}_{1}$ -- 3s\,$^{1}$S$_{0}$                          &    0.55  &  0.46$\pm$0.12  &    ...   &  ...            &    0.62  &  0.52$\pm$0.13  &    0.55  &  0.44$\pm$0.11\\
$[$O~{\sc ii}$]$   &  7320       & 2p$^{3}$\,$^{2}$D$^{\rm o}_{5/2}$ -- 2p$^{3}$\,$^{2}$P$^{\rm o}_{3/2}$  &    8.40  &  7.02$\pm$1.14  &    9.79  &  8.17$\pm$1.22  &   10.4   &  8.82$\pm$1.32  &    5.62  &  4.49$\pm$0.67\\
$[$O~{\sc ii}$]$   &  7330$^{m}$ & 2p$^{3}$\,$^{2}$D$^{\rm o}_{3/2}$ -- 2p$^{3}$\,$^{2}$P$^{\rm o}_{3/2}$  &    8.08  &  6.74$\pm$1.00  &   10.9   &  9.13$\pm$1.37  &    7.13  &  6.06$\pm$0.90  &    4.21  &  3.36$\pm$0.50\\
$[$Ar~{\sc iii}$]$ &  7751       & 3p$^{4}$\,$^{3}$P$_{1}$ -- 3p$^{4}$\,$^{1}$D$_{2}$                      &    3.86  &  3.14$\pm$1.16  &    1.67  &  1.35$\pm$0.47  &    7.28  &  6.05$\pm$2.45  &    5.31  &  4.10$\pm$1.44\\
\\
$c$(H$\beta$)      &       &                                                                       & \multicolumn{2}{c}{0.196}    & \multicolumn{2}{c}{0.198}    & \multicolumn{2}{c}{0.177}    & \multicolumn{2}{c}{0.247}   \\
\multicolumn{2}{l}{$\log$\,$F$(H$\beta$)$^{n}$}  &                                                 & \multicolumn{2}{c}{$-$14.70} & \multicolumn{2}{c}{$-$14.96} & \multicolumn{2}{c}{$-$14.73} & \multicolumn{2}{c}{$-$14.88}\\
\rule[-1mm]{0mm}{5.0mm}\\
\hline
\rule[-1mm]{0mm}{5.0mm}\\
Ion  & $\lambda$  & Transition &
\multicolumn{2}{c}{\underline{~~~~~~~PN5~~~~~~~}} &
\multicolumn{2}{c}{\underline{~~~~~~~PN6~~~~~~~}} &
\multicolumn{2}{c}{\underline{~~~~~~~PN7~~~~~~~}} \\
     &  (\AA)     &            &
$F$($\lambda$) & $I$($\lambda$) &
$F$($\lambda$) & $I$($\lambda$) &
$F$($\lambda$) & $I$($\lambda$) \\
\hline
\rule[-1mm]{0mm}{5.0mm}\\
$[$O~{\sc ii}$]$   &  3727$^{b}$ & 2p$^{3}$\,$^{4}$S$^{\rm o}$ -- 2p$^{3}$\,$^{2}$D$^{\rm o}$              &    42.4   &  59.4$\pm$7.1  &   33.2   &  46.4$\pm$9.3  &   73.4   &  84.3$\pm$7.6 \\
H~{\sc i}          &  3798       & 2p\,$^{2}$P$^{\rm o}$ -- 10d\,$^{2}$D                                   &     ...   &   ...          &   ...    &   ...          &    5.23  &  5.97$\pm$1.64\\
H~{\sc i}          &  3835       & 2p\,$^{2}$P$^{\rm o}$ -- 9d\,$^{2}$D                                    &     ...   &   ...          &   10.2   &  13.9$\pm$3.5  &    4.17  &  4.74$\pm$1.20\\
$[$Ne~{\sc iii}$]$ &  3868       & 2p$^{4}$\,$^{3}$P$_{2}$ -- 2p$^{4}$\,$^{1}$D$_{2}$                      &    87.0   &   118$\pm$6    &   63.0   &  85.4$\pm$4.2  &   75.6   &  85.6$\pm$4.2 \\
H~{\sc i}          &  3889$^{c}$ & 2p\,$^{2}$P$^{\rm o}$ -- 8d\,$^{2}$D                                    &    13.0   &  17.6$\pm$2.0  &   16.0   &  21.6$\pm$2.3  &   18.3   &  20.8$\pm$2.3 \\
\end{tabular}
\end{center}
\end{table*}

\addtocounter{table}{-1}
\begin{table*}
\begin{center}
\caption{(Continued)}
\label{lines}
\begin{tabular}{lllcccccccc}
\hline\hline
\rule[-1mm]{0mm}{5.0mm}\\
Ion  & $\lambda$  & Transition & 
\multicolumn{2}{c}{\underline{~~~~~~~PN5~~~~~~~}} & 
\multicolumn{2}{c}{\underline{~~~~~~~PN6~~~~~~~}} & 
\multicolumn{2}{c}{\underline{~~~~~~~PN7~~~~~~~}} & 
\multicolumn{2}{c}{~~~~~~~~~~~~~~~~~~~~~~~}\\
     &  (\AA)     &            & 
$F$($\lambda$) & $I$($\lambda$) & 
$F$($\lambda$) & $I$($\lambda$) & 
$F$($\lambda$) & $I$($\lambda$) & & \\
\hline
\rule[-1mm]{0mm}{5.0mm}\\
$[$Ne~{\sc iii}$]$ &  3967$^{d}$ & 2p$^{4}$\,$^{3}$P$_{1}$ -- 2p$^{4}$\,$^{1}$D$_{2}$                      &    43.2   &  57.0$\pm$5.0  &   32.3   &  42.7$\pm$3.3  &   42.2   &  47.3$\pm$3.7 \\
$[$S~{\sc ii}$]$   &  4068$^{e}$ & 3p$^{3}$\,$^{4}$S$^{\rm o}_{3/2}$ -- 3p$^{3}$\,$^{2}$P$^{\rm o}_{3/2}$  &     6.36  &   8.2$\pm$3.2  &    ...   &   ...          &    4.37  &  4.84$\pm$1.16\\
H~{\sc i}          &  4101       & 2p\,$^{2}$P$^{\rm o}$ -- 6d\,$^{2}$D                                    &    27.8   &  35.0$\pm$4.1  &   30.4   &  38.7$\pm$4.3  &   25.5   &  28.2$\pm$3.1 \\
H~{\sc i}          &  4340$^{f}$ & 2p\,$^{2}$P$^{\rm o}$ -- 5d\,$^{2}$D                                    &    37.8   &  44.7$\pm$2.0  &   38.8   &  45.8$\pm$2.1  &   44.0   &  47.1$\pm$2.1 \\
$[$O~{\sc iii}$]$  &  4363       & 2p$^{2}$\,$^{1}$D$_{2}$ -- 2p$^{2}$\,$^{1}$S$_{0}$                      &     8.22  &   9.6$\pm$1.2  &    4.84  &  5.68$\pm$0.70 &    6.79  &  7.25$\pm$0.80\\
He~{\sc i}         &  4471       & 2p\,$^{3}$P$^{\rm o}$ -- 4d\,$^{3}$D                                    &     7.82  &  8.84$\pm$1.40 &    4.50  &  5.10$\pm$0.80 &    6.54  &  6.87$\pm$0.76\\
N~{\sc iii}        &  4641       & 3p\,$^{2}$P$^{\rm o}_{3/2}$ -- 3d\,$^{2}$D$_{5/2}$                      &     ...   &   ...          &    ...   &   ...          &    2.10  &  2.15$\pm$0.43\\
He~{\sc ii}        &  4686       & 3d\,$^{2}$D -- 4f\,$^{2}$F$^{\rm o}$                                    &     1.85  &  1.95$\pm$0.30 &    1.76  &  1.85$\pm$:    &    2.23  &  2.28$\pm$0.34\\
$[$Ar~{\sc iv}$]$  &  4711$^{i}$ & 3p$^{3}$\,$^{4}$S$^{\rm o}_{3/2}$ -- 3p$^{3}$\,$^{2}$D$^{\rm o}_{5/2}$  &     2.23  &  2.33$\pm$0.28 &    ...   &   ...          &    1.55  &  1.58$\pm$0.20\\
$[$Ar~{\sc iv}$]$  &  4740       & 3p$^{3}$\,$^{4}$S$^{\rm o}_{3/2}$ -- 3p$^{3}$\,$^{2}$D$^{\rm o}_{3/2}$  &     3.30  &  3.43$\pm$0.40 &    ...   &   ...          &    1.67  &  1.70$\pm$0.22\\
H~{\sc i}          &  4861       & 2p\,$^{2}$P$^{\rm o}$ -- 4d\,$^{2}$D                                    &   100     &   100          &  100     &   100          &  100     &   100         \\
He~{\sc i}         &  4922       & 2p\,$^{1}$P$^{\rm o}_{1}$ -- 4d\,$^{1}$D$_{2}$                          &     3.03  &  3.00$\pm$1.24 &    ...   &   ...          &    2.13  &  2.12$\pm$0.84\\
$[$O~{\sc iii}$]$  &  4959       & 2p$^{2}$\,$^{1}$P$_{1}$ -- 2p$^{2}$\,$^{1}$D$_{2}$                      &   422     &   413$\pm$20   &  376     &   368$\pm$18   &  342     &   339$\pm$16  \\
$[$O~{\sc iii}$]$  &  5007       & 2p$^{2}$\,$^{1}$P$_{2}$ -- 2p$^{2}$\,$^{1}$D$_{2}$                      &  1260     &  1216$\pm$37   & 1106     &  1068$\pm$32   & 1033     &  1017$\pm$31  \\
$[$N~{\sc i}$]$    &  5198$^{j}$ & 2p$^{3}$\,$^{4}$S$^{\rm o}_{3/2}$ -- 2p$^{3}$\,$^{2}$D$^{\rm o}_{3/2}$  &     ...   &   ...          &   ...    &   ...          &    1.62  &  1.57$\pm$0.55\\
$[$N~{\sc ii}$]$   &  5755       & 2p$^{2}$\,$^{1}$D$_{2}$ -- 2p$^{2}$\,$^{1}$S$_{0}$                      &     2.57  &  2.12$\pm$0.64 &   ...    &   ...          &    1.10  &  1.02$\pm$0.30\\
He~{\sc i}         &  5876       & 2p\,$^{3}$P$^{\rm o}$ -- 3d\,$^{3}$D                                    &    21.2   &  17.2$\pm$1.60 &   16.4   &  13.3$\pm$1.5  &   17.9   &  16.4$\pm$0.8 \\
$[$O~{\sc i}$]$    &  6300       & 2p$^{4}$\,$^{3}$P$_{2}$ -- 2p$^{4}$\,$^{1}$D$_{2}$                      &     6.87  &  5.24$\pm$1.30 &   ...    &   ...          &    7.88  &  7.04$\pm$1.76\\
$[$S~{\sc iii}$]$  &  6312$^{k}$ & 3p$^{2}$\,$^{1}$D$_{2}$ -- 3p$^{2}$\,$^{1}$S$_{0}$                      &     ...   &   ...          &   ...    &   ...          &    2.03  &  1.81$\pm$0.56\\
$[$O~{\sc i}$]$    &  6363       & 2p$^{4}$\,$^{3}$P$_{1}$ -- 2p$^{4}$\,$^{1}$D$_{2}$                      &     3.73  &  2.82$\pm$1.26 &   ...    &   ...          &    2.33  &  2.08$\pm$0.80\\
$[$N~{\sc ii}$]$   &  6548       & 2p$^{2}$\,$^{3}$P$_{1}$ -- 2p$^{2}$\,$^{1}$D$_{2}$                      &    33.1   &  24.4$\pm$2.9  &    5.91  &  4.36$\pm$0.52 &   28.1   &  24.8$\pm$3.0 \\
H~{\sc i}          &  6563       & 2p\,$^{2}$P$^{\rm o}$ -- 2d\,$^{2}$D                                    &   357     &   263$\pm$1    &  338     &   249$\pm$1    &  341     &   300$\pm$1   \\
$[$N~{\sc ii}$]$   &  6583       & 2p$^{2}$\,$^{3}$P$_{2}$ -- 2p$^{2}$\,$^{1}$D$_{2}$                      &    96.5   &  71.0$\pm$7.5  &   18.0   &  13.2$\pm$1.5  &   87.3   &  76.8$\pm$8.4 \\
He~{\sc i}         &  6678$^{l}$ & 2p\,$^{1}$P$^{\rm o}_{1}$ -- 3d\,$^{1}$D$_{2}$                          &     5.39  &  3.90$\pm$0.58 &    ...   &   ...          &    5.11  &  4.47$\pm$0.67\\
$[$S~{\sc ii}$]$   &  6716       & 3p$^{3}$\,$^{4}$S$^{\rm o}_{3/2}$ -- 3p$^{3}$\,$^{2}$D$^{\rm o}_{5/2}$  &    10.3   &  7.45$\pm$0.83 &    ...   &   ...          &   10.1   &   8.8$\pm$1.0 \\
$[$S~{\sc ii}$]$   &  6731       & 3p$^{3}$\,$^{4}$S$^{\rm o}_{3/2}$ -- 3p$^{3}$\,$^{2}$D$^{\rm o}_{3/2}$  &    13.6   &   9.8$\pm$1.0  &    ...   &   ...          &   15.4   &  13.5$\pm$1.0 \\
$[$Ar~{\sc v}$]$   &  7005       & 3p$^{2}$\,$^{3}$P$_{2}$ -- 3p$^{2}$\,$^{1}$D$_{2}$                      &     ...   &   ...          &    ...   &   ...          &    0.50  &  0.42$\pm$0.11\\ 
He~{\sc i}         &  7065       & 2p\,$^{3}$P$^{\rm o}$ -- 3s\,$^{3}$S                                    &     9.90  &  6.80$\pm$0.86 &    ...   &   ...          &   11.6   &  10.0$\pm$1.3 \\
$[$Ar~{\sc iii}$]$ &  7136       & 3p$^{4}$\,$^{3}$P$_{2}$ -- 3p$^{4}$\,$^{1}$D$_{2}$                      &    34.5   &  23.4$\pm$3.0  &    ...   &   ...          &   20.9   &  17.8$\pm$2.13\\
$[$O~{\sc ii}$]$   &  7320       & 2p$^{3}$\,$^{2}$D$^{\rm o}_{5/2}$ -- 2p$^{3}$\,$^{2}$P$^{\rm o}_{3/2}$  &     ...   &   ...          &    ...   &   ...          &    5.77  &  4.87$\pm$0.73\\
$[$O~{\sc ii}$]$   &  7330$^{m}$ & 2p$^{3}$\,$^{2}$D$^{\rm o}_{3/2}$ -- 2p$^{3}$\,$^{2}$P$^{\rm o}_{3/2}$  &     ...   &   ...          &    ...   &   ...          &    5.94  &  5.01$\pm$0.75\\
$[$Ar~{\sc iii}$]$ &  7751       & 3p$^{4}$\,$^{3}$P$_{1}$ -- 3p$^{4}$\,$^{1}$D$_{2}$                      &    11.1   &  6.93$\pm$2.43 &    ...   &   ...          &    4.29  &  3.53$\pm$1.24\\
\\
$c$(H$\beta$)      &       &                                                                       & \multicolumn{2}{c}{0.451}    & \multicolumn{2}{c}{0.450}    & \multicolumn{2}{c}{0.186}   \\
\multicolumn{2}{l}{$\log$\,$F$(H$\beta$)$^{n}$}  &                                                 & \multicolumn{2}{c}{$-$14.94} & \multicolumn{2}{c}{$-$15.13} & \multicolumn{2}{c}{$-$15.10}\\
\rule[-1mm]{0mm}{5.0mm}\\
\hline
\end{tabular}
\begin{description}
\item[$^{a}$] Fluxes (also the intensities) are normalized such that 
H$\beta$ = 100.  Emission lines that were not detected in the spectra of 
PN5, PN6 or PN7 are not presented in the table for the three targets.

\item[$^{b}$] A blend of the O~{\sc ii} $\lambda$3726 
(2p$^{3}$\,$^{4}$S$^{\rm o}_{3/2}$ -- 2p$^{3}$\,$^{2}$D$^{\rm o}_{3/2}$) and 
$\lambda$3729 (2p$^{3}$\,$^{4}$S$^{\rm o}_{3/2}$ -- 2p$^{3}$\,$^{2}$D$^{\rm 
o}_{5/2}$) lines.

\item[$^{c}$] Blended with the He~{\sc i} $\lambda$3888 
(2s\,$^{3}$S -- 3p\,$^{3}$P$^{\rm o}$) line.

\item[$^{d}$] Blended with the H~{\sc i} $\lambda$3970 
(2p\,$^{2}$P$^{\rm o}$ -- 7d\,$^{2}$D) and He~{\sc i} $\lambda$3965 
(2s\,$^{1}$S$_{0}$ -- 4p\,$^{1}$P$^{\rm o}_{1}$) lines.

\item[$^{e}$] Blended with the O~{\sc ii} M10 3p\,$^{4}$D$^{\rm o}$ -- 
3d\,$^{4}$F and C~{\sc iii} M16 4f\,$^{3}$F$^{\rm o}$ -- 5g\,$^{3}$G, 
and [S~{\sc ii}] 4076 (3p$^{3}$\,$^{4}$S$^{\rm o}_{3/2}$ -- 
3p$^{3}$\,$^{2}$P$^{\rm o}_{1/2}$) lines.

\item[$^{f}$] Corrected for the flux contribution from the blended He~{\sc 
ii} $\lambda$4338 (4f\,$^{2}$F$^{\rm o}$ -- 10g\,$^{2}$G) line.  Fluxes of 
H$\beta$ and H$\alpha$ have also been corrected for the blended He~{\sc ii} 
$\lambda$4859 (4f\,$^{2}$F$^{\rm o}$ -- 8g\,$^{2}$G) and $\lambda$6560 
(4f\,$^{2}$F$^{\rm o}$ -- 6g\,$^{2}$G) lines, respectively.

\item[$^{g}$] Blended with the N~{\sc iii} $\lambda$4634 and $\lambda$4642 
lines of multiplet M2 3p\,2P$^{\rm o}$ -- 3d\,$^{2}$D and the O~{\sc ii} 
$\lambda$4642 and $\lambda$4639 lines of multiplet M1 3s\,$^{4}$P -- 
3p\,$^{4}$D$^{\rm o}$.

\item[$^{h}$] Probably blended with the C~{\sc iv} $\lambda$4658 (M8 
5g\,$^{2}$G -- 6h\,$^{2}$H$^{\rm o}$) line for the case of PN2.

\item[$^{i}$] Corrected for the flux contribution from the blended He~{\sc 
i} $\lambda$4713 (2p\,$^{3}$P$^{\rm o}$ -- 4s\,$^{3}$S) line.

\item[$^{j}$] Blended with [N~{\sc i}] $\lambda$5200 
(2p$^{3}$\,$^{4}$S$^{\rm o}_{3/2}$ -- 2p$^{3}$\,$^{2}$D$^{\rm o}_{5/2}$.

\item[$^{k}$] Corrected for the flux contribution from the blended He~{\sc 
ii} $\lambda$6311 (5g\,$^{2}$G -- 16h\,$^{2}$H$^{\rm o}$) line.

\item[$^{l}$] Corrected for the flux contribution from the blended He~{\sc 
ii} $\lambda$6683 (5g\,$^{2}$G -- 13h\,$^{2}$H$^{\rm o}$) line. 

\item[$^{m}$] Blended with the [Ar~{\sc iv}] $\lambda$7331 (3p$^{3}$ 
$^{2}$D$^{\rm o}_{5/2}$ -- $^{2}$P$^{\rm o}_{1/2}$) line, whose flux 
contribution is negligible ($<$1\%).

\item[$^{n}$] In unit of erg\,cm$^{-2}$\,s$^{-1}$ in the extracted spectra.
\end{description}
\end{center}
\end{table*}

\subsection{Plasma Diagnostics} \label{section3:c}

Plasma diagnostics were determined using the collisionally 
excited lines\footnote{Quite often, those lines are electron 
dipole-forbidden transitions, so they are also commonly called 
forbidden lines.} 
(CELs) of heavy elements in conventional manner \citep[e.g.,][]{of06}.  
The [O~{\sc iii}] ($\lambda$4959 + $\lambda$5007)/$\lambda$4363 
auroral-to-nebular line ratio was used to derive the electron temperature. 
The [N~{\sc ii}] $\lambda$5755 auroral line, whose flux is typically 
$\sim$5--15\% of that of the [O~{\sc iii}] $\lambda$4363 line, was 
detected in all targets except PN6, and the [N~{\sc ii}] ($\lambda$6548 
+ $\lambda$6583)/$\lambda$5755 ratio was also used as a temperature 
diagnostic.  The [Ar~{\sc iv}] $\lambda$4711/$\lambda$4740 and [S~{\sc 
ii}] $\lambda$6716/$\lambda$6731 nebular line ratios were used for 
density diagnostics.  The [O~{\sc ii}] $\lambda$3727 line is a blend 
of the $\lambda\lambda$3726,\,3729 doublet.  Since the [O~{\sc ii}] 
$\lambda\lambda$7320,\,7330 auroral lines were detected in the spectra 
of PN1--4 and PN7, we also used the [O~{\sc ii}] 
$\lambda$3727/($\lambda$7320 + $\lambda$7330) line ratio to estimate 
electron temperatures for these five objects. 
The weak [S~{\sc ii}] $\lambda$4068 line is blended with the 
even weaker [S~{\sc ii}] $\lambda$4076 line as well as with the O~{\sc 
ii} M10 3p\,$^{4}$D$^{\rm o}$ -- 3d\,$^{4}$F and C~{\sc iii} M16 
4f\,$^{3}$F$^{\rm o}$ -- 5g\,$^{3}$G lines.  The electron temperatures 
derived from the [S~{\sc ii}] ($\lambda$6716 + $\lambda$6731)/$\lambda$4072 
ratio are highly uncertain.

\begin{figure*}
\begin{center}
\includegraphics[width=11.5cm,angle=-90]{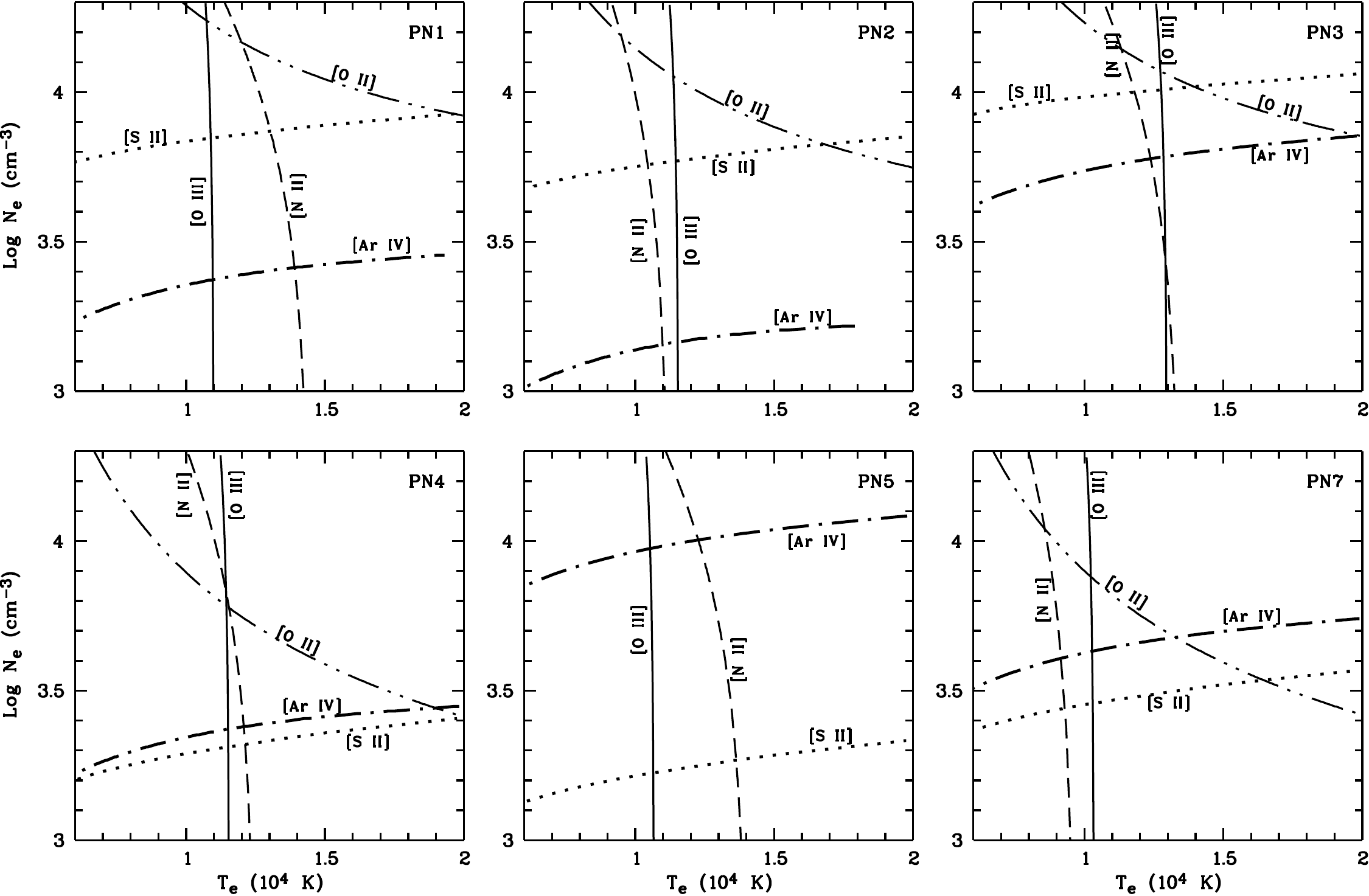}
\caption{
Plasma diagnostic diagrams for PN1 (top left), PN2 (top middle), PN3 (top 
right), PN4 (bottom left), PN5 (bottom middle) and PN7 (bottom right). 
Different line types represent the temperature or density diagnostics 
using different ions (see Table~\ref{temden} for line ratios). 
The dotted line is the [S~{\sc ii}] $\lambda$6716/$\lambda$6731 
density-diagnostic curve. 
The diagram of PN6 is not presented due to the lack of diagnostics (see 
text for details). } 
\label{diagnose}
\end{center}
\end{figure*}

Figure~\ref{diagnose} shows the plasma diagnostic diagrams for six of 
our PNe.  The diagram of PN6 is not presented because only the [O~{\sc 
iii}] $\lambda$4363 diagnostic line was detected in the spectrum of 
this object. 
These diagnostic diagrams are based on CELs and were created by 
solving the level population equations for five-level atomic models 
using {\sc equib}, a {\sc fortran} code originally developed by 
\citet{ha81} to solve the equations of statistical equilibrium in 
multi-level atoms and yield level populations and line emissivities 
for a specified physical condition, appropriate to the zones where 
the ions are expected to exist.  The atomic data used by the {\sc 
equib} code were updated manually.  References for the atomic 
data set used for plasma diagnostics based on CELs, as well as for 
ionic abundance determinations (Section~\ref{section3:d:1}), are 
presented in Table~\ref{atomic_data}.  Results of our plasma diagnostics 
are presented in Table~\ref{temden}.

\begin{table}
\begin{center}
\begin{minipage}{90mm}
\caption{References for Atomic Data}
\label{atomic_data}
\begin{tabular}{lll}
\hline
\hline
Ion & \multicolumn{2}{c}{CELs}\\
    & Transition probabilities & Collision strengths\\
\hline
N$^{+}$   & \citet{bhs95} & \citet{sta94}\\
O$^{+}$   & \citet{zeip87}  & \citet{pra06}\\
O$^{2+}$  & \citet{sz00}  & \citet{lb94}\\
Ne$^{2+}$ & \citet{lb05} & \citet{mb00}\\
S$^{+}$   & \citet{kee93} & \citet{ram96}\\
S$^{2+}$  & \citet{mz82a} & \citet{tg99}\\
Ar$^{2+}$ & \citet{bh86} & \citet{gmz95}\\
Ar$^{3+}$ & \citet{mz82b} & \citet{ram97}\\
Ar$^{4+}$ & \citet{mz82a} & \citet{med83}\\
\hline
Ion & \multicolumn{2}{c}{ORLs}\\
    & Effective recombination coeff. & Comments\\
\hline
H~{\sc i}   & \citet{sh95}     & Case~B\\
He~{\sc i}  & \citet{porter12} & Case~B\\
He~{\sc ii} & \citet{sh95}     & Case~B\\
C~{\sc ii}  & \citet{dav00}    & Case~B\\
\hline
\end{tabular}
\end{minipage}
\end{center}
\end{table}

The uncertainties in the electron temperatures and densities 
(Table~\ref{temden}) were estimated through the propagation of 
uncertainties in the emission line intensities, and do not include 
the contribution from systematic effects.  As discussed in 
Section~\ref{section3:a}, the error contributions from the systematic 
effects, such as reddening, flux calibration, and line blending, are 
expected to be small.  The atomic data used for plasma diagnostics are 
mostly updated (Table~\ref{atomic_data}).  It is difficult to 
quantitatively estimate the systematic uncertainties introduced by 
the atomic data.  We expect that the atomic data adopted are reliable, 
since these calculations were improved over the previous work (see 
the discussion in Section~7 of \citealt{kwi14} contributed by X.-W. Liu 
and X. Fang).

Figure~\ref{diagnose} shows that the [O~{\sc iii}] line ratio is an 
excellent temperature diagnostic, and variation of the electron density 
does not change the resultant electron temperature much. 
Uncertainties in the electron temperatures derived from the [N~{\sc 
ii}] line ratio are much larger than those of the [O~{\sc iii}] 
temperatures, mainly due to weakness of the [N~{\sc ii}] $\lambda$5755 
auroral line. 
Besides, the [N~{\sc ii}] nebular-to-auroral line ratio has also 
slight density dependence (Figure~\ref{diagnose}).  The [O~{\sc ii}] 
nebular-to-auroral line ratio is sensitive to both temperature and 
density, and thus not an ideal temperature indicator. 
For some targets, the [N~{\sc ii}] temperature is higher than 
the [O~{\sc iii}] temperature, which is likely due to the overestimated 
intensity of the very weak $\lambda$5755 line.  Only in one object (PN7) 
the [N~{\sc ii}] temperature seems to be meaningfully different from 
the [O~{\sc iii}] temperature (Table~\ref{temden}). 
The [O~{\sc iii}] temperature is thus adopted in the abundance 
calculations for all PNe. 
We are aware that in a PN there are different ionization zones, as 
represented by the O$^{2+}$ and N$^{+}$ ions, and using only the [O~{\sc 
iii}] temperature may introduce extra errors in the ionic abundances for 
the singly ionized species.  In this respect, the actual uncertainties 
in the N$^{+}$/H$^{+}$, O$^{+}$/H$^{+}$ and S$^{+}$/H$^{+}$ abundance 
ratios might be slightly larger than those given in Table~\ref{ionic}. 
Recently, in a spectroscopic study of Galactic PNe, \citet{duf15} 
adopted a constant [N~{\sc ii}] temperature, which was carefully 
derived by \citet{kaler86} for the case where there is no detection of 
the [N~{\sc ii}] $\lambda$5755 line but He~{\sc ii} $\lambda$4686 is 
present.

In a few cases of our PN sample, there is large difference between the 
electron density derived from the [S~{\sc ii}] line ratio and that 
derived from [Ar~{\sc iv}], as shown in Table~\ref{temden}. 
It might be that in our PNe there are low-ionization regions where 
the [S~{\sc ii}] density is more appropriate, whereas the [Ar~{\sc iv}] 
density applies in the higher ionization regions.  Thus we adopted the 
[S~{\sc ii}] densities for the ionic-abundance calculations of the low 
ionization species and adopted the [Ar~{\sc iv}] densities for the high 
ionization species (Section~\ref{section3:d:1}). 
Measurements of the relatively faint [Ar~{\sc iv}] lines in PN5 are of 
large uncertainty due to the noise, and for this object we adopted the 
[S~{\sc ii}] density.

An electron density of $\sim$2000~cm$^{-3}$, typical for most of our 
targets, was assumed for PN6 because neither the [S~{\sc ii}] nor the 
[Ar~{\sc iv}] doublet was detected in this PN.  In order to check how much 
the assumed electron density would affect the resultant ionic abundances, 
we calculated ionic abundances for two density cases, 10$^{3}$ and 
10$^{4}$~cm$^{-3}$, at 9500~K, the [O~{\sc iii}] temperature derived for 
PN6.  The N$^{+}$/H$^{+}$ ionic abundance ratio derived at the two 
density cases differ by $\sim$10\%, while both the O$^{2+}$/H$^{+}$ 
and the Ne$^{2+}$/H$^{+}$ ratio differ by $<$4\%; the O$^{+}$/H$^{+}$ 
abundance ratio derived from the [O~{\sc ii}] $\lambda$3727 line differ
by nearly 50\% at the two densities. 
These differences in the CEL ionic abundances at the two density cases 
are due to the large differences in critical densities, as discussed 
in Paper~I.  For the helium recombination lines, differences in ionic 
abundances derived at the two density cases are negligible ($<$6\%; 
\citealt{porter12}).

The relatively high electron temperature and density of PN3 
(Table~\ref{temden}) indicate that this target may be young (compact) 
compared to other PNe.  The [O~{\sc iii}] $\lambda$4363 auroral line 
in PN3 is stronger than other PNe.  PN3 also has strong He~{\sc ii} 
$\lambda$4686 line ($\sim$0.2 H$\beta$), while this line is much fainter 
in the spectra of the other targets (Table~\ref{lines}).  The above 
information points to the high-excitation nature of PN3.  This high 
excitation could due to either a very hot central star or the low 
metallicity, or both.  The relatively high density in PN3 may also cause 
collisional de-excitation of the heavy element ions, which makes the 
radiation cooling inefficient and as a consequence, produces relatively 
higher electron temperature. 

The intensity ratio of two He~{\sc i} optical recombination lines that 
originate from two atomic levels with different excitation energies is 
sensitive to the electron temperature (although not so acutely sensitive 
to $T_\mathrm{e}$ as the CELs), and thus can also be used as a temperature 
diagnostic.  The temperature diagnostics for gaseous nebulae based on the 
He~{\sc i} recombination lines were constructed by \citet{zhang05} using 
the theoretical He~{\sc i} line emissivities calculated by \citet{bss99}.
We have determined the He~{\sc i} line temperatures for our PNe 
following the diagnostic method of \citet{zhang05} and the results are 
presented in Table~\ref{temden}.  Generally the electron temperatures 
derived from the He~{\sc i} line ratios, designated $T_\mathrm{e}$(He~{\sc 
i}), are lower than those derived from the heavy element CELs, 
$T_\mathrm{e}$(CEL), except PN4 and PN7, whose He~{\sc i} 
$\lambda$6678/$\lambda$4471 ratio yields higher temperatures.  This is in 
line with the temperature sequence, $T_\mathrm{e}$(CEL) $\gtrsim$ 
$T_\mathrm{e}$(H~{\sc i}) $\gtrsim$ $T_\mathrm{e}$(He~{\sc i}), so far 
observed in more than 100 Galactic PNe through deep spectroscopy (e.g., 
\citealt{mcn13}; see also recent reviews by \citealt{liu06,liu12}).  
\citet{zhang05} studied 48 Galactic PNe and found that the 
$T_\mathrm{e}$(He~{\sc i}) values are significantly lower than the electron
temperatures derived from the H~{\sc i} Balmer jump, $T_\mathrm{e}$(H~{\sc 
i}), with an average difference of 4000~K.  The significant difference 
between the electron temperatures of PNe derived from the CELs and the 
H~{\sc i} (also He~{\sc i}) recombination spectrum is the renowned 
``temperature discrepancy'' in nebular astrophysics \citep[e.g.,][]{peim71,
ld93}.  Investigation of this subject has never been extended to 
extragalactic PNe due to limited S/N.  The spectral quality of our PN 
targets also prevented further study of the problem.

\begin{table*}
\begin{center}
\caption{Plasma Diagnostics}
\label{temden}
\begin{tabular}{lccccccc}
\hline
\hline
\rule[-0.5mm]{0mm}{1.0mm}\\
Diagnostic Ratio & PN1 & PN2 & PN3 & PN4 & PN5 & PN6 & PN7\\
\hline
                 & \multicolumn{7}{c}{$T_\mathrm{e}$ (K)} \\
$[$O~{\sc iii}$]$ ($\lambda$4959+$\lambda$5007)/$\lambda$4363  &  11\,000$\pm$400 &  11\,600$\pm$500 &  12\,900$\pm$550 &  11\,600$\pm$700 &  10\,600$\pm$700 &  9500$\pm$600 & 10\,300$\pm$350\\
$[$N~{\sc ii}$]$  ($\lambda$6548+$\lambda$6583)/$\lambda$5755  & 14\,300$\pm$2400 & 11\,300$\pm$1500 & 12\,300$\pm$1300 & 12\,400$\pm$2000 & 11\,800$\pm$2000 &  ...          &   9100$\pm$1000\\
$[$O~{\sc ii}$]$  $\lambda$3727/($\lambda$7320+$\lambda$7330)  &  $>$20\,000      &   $>$20\,000     & 15\,200$\pm$2200 & 14\,600$\pm$3000 &  ...             &  ...          & 11\,000$\pm$1100\\
$[$S~{\sc ii}$]$  ($\lambda$6716+$\lambda$6731)/$\lambda$4072$^{a}$  & ...        & ...              & 12\,000$\pm$6700 & 15\,400$\pm$8000 &  ...             &  ...          & 10\,000$\pm$8600\\
He~{\sc i} $\lambda$5876/$\lambda$4471                         &   5300$\pm$2000  &    6100$\pm$2000 &    8000$\pm$2000 &   $<$5000        &  ...             &  ...          & $<$5000         \\
He~{\sc i} $\lambda$6678/$\lambda$4471                         &   5200$\pm$2500  &    3400$\pm$3000 &    9700$\pm$3000 &   12500$\pm$7500 &  ...             &  ...          &   15000$\pm$6000\\
Adopted$^{b}$                                                  &  11\,000$\pm$400 &  11\,600$\pm$500 &  12\,900$\pm$550 &  11\,600$\pm$500 &  10\,600$\pm$400 &  9500$\pm$400 & 10\,300$\pm$350\\
\\
                  & \multicolumn{7}{c}{$N_\mathrm{e}$ (cm$^{-3}$)}\\
$[$S~{\sc ii}$]$  $\lambda$6716/$\lambda$6731 &  7000$\pm$2300 &  5900$\pm$2100 & 10\,200$\pm$4400 &  2000$\pm$1000 &  1700$\pm$800  & 2000$^{c}$ &  2900$\pm$1200\\
$[$Ar~{\sc iv}$]$ $\lambda$4711/$\lambda$4740 &  3000$\pm$200  &  1900$\pm$300  &    7700$\pm$200  &  2900$\pm$400  &  9400$\pm$2000 & ...        &  5400$\pm$600\\
\hline
\end{tabular}
\begin{description}
\item[$^{a}$] A blend of [S~{\sc ii}] $\lambda\lambda$4068,\,4076 and 
the O~{\sc ii} M10 3p\,$^{4}$D$^{\rm o}$ -- 3d\,$^{4}$F and C~{\sc iii} 
M16 4f\,$^{3}$F$^{\rm o}$ -- 5g\,$^{3}$G lines.

\item[$^{b}$] The electron temperatures derived from the [O~{\sc iii}] line 
ratio are adopted.


\item[$^{c}$] An assumed electron density for PN6 because neither [S~{\sc 
ii}] nor [Ar~{\sc iv}] lines were observed.
\end{description}
\end{center}
\end{table*}

\subsection{Abundances} \label{section3:d}

\subsubsection{Ionic Abundances} \label{section3:d:1}

In order to derive ionic abundances of the heavy elements from CELs, 
the equations of statistical equilibrium were solved using the program 
{\sc equib} to determine the population of the upper level of a given 
transition.  References of the atomic data used for ionic abundance 
calculations are summarized in Table~\ref{atomic_data}.  The electron 
temperatures and densities (Table~\ref{temden}) obtained from plasma 
diagnostics (Section~\ref{section3:c}) were assumed in the abundance 
calculations.  Ionic abundances derived using the extinction-corrected 
fluxes of the lines detected in the spectra of the seven PNe are 
presented in Table~\ref{ionic}.
\\

\noindent
{\it Helium} -- 
The He$^{+}$/H$^{+}$ ionic abundances were derived from the He~{\sc i} 
$\lambda\lambda$3888, 4471, 5876 and 6678 lines (Table~\ref{ionic}), which
are among the strongest helium recombination lines in the optical.  The 
effective recombination coefficients of the He~{\sc i} lines calculated 
by \citet{porter12} were adopted in the abundance determination. 
The He~{\sc i} $\lambda$3888 line is blended with H~{\sc i} $\lambda$3889 
($n$ = 2--8), which contributes $\sim$50--60\% of the total flux.  We 
corrected for the blended H~{\sc i} line flux using the theoretical 
H~{\sc i} $\lambda$3889/$\lambda$4861 line ratio at 10\,000~K and 
10$^{4}$~cm$^{-3}$ in Case~B \citep{sh95}.  
However, the ionic abundance yielded by the corrected $\lambda$3888 line 
flux is lower than those derived from the $\lambda\lambda$4471, 5876 and 
6678 lines for our targets (Table~\ref{ionic}).  Such significant 
difference between the He$^{+}$/H$^{+}$ abundance derived from the 
$\lambda$3888 line and those derived from the other three He~{\sc i} lines 
is difficult to explain.  One possibility is the effect of self-absorption 
due to large population in the 1s2s\,$^{3}$S$_{1}$ metastable level of 
He~{\sc i} \citep[e.g.,][]{rob68,bss02}.  The limited number of 
well-detected He~{\sc i} lines in our sample prevents further study of 
such effect.  The only agreement in the He$^{+}$/H$^{+}$ abundances in 
found in PN6.  The He$^{+}$/H$^{+}$ abundance derived from the He~{\sc i} 
$\lambda$5876 line is adopted for our sample, given that this line is 
stronger than the other He~{\sc i} optical recombination lines and thus 
best observed.


The He$^{2+}$/H$^{+}$ abundance was derived from the He~{\sc ii} 
$\lambda$4686 line, which was detected in the spectra of all targets except 
PN6, for which an upper limit of the line flux was estimated from the 
standard deviation of the local continuum.  This upper limit value is 
close to the 
He~{\sc ii} line fluxes of other faint PNe.  Another He~{\sc ii} line 
$\lambda$5411 was only detected in PN3, which also has the strongest 
$\lambda$4686 line.  The He$^{2+}$/H$^{+}$ abundance of PN3 is higher 
than those of the other PNe by nearly one order of magnitude, indicating 
that this PN is highly excited.  The effective recombination coefficients 
of the He~{\sc ii} line used in the abundance determination was adopted 
from the hydrogenic calculations of \citet{sh95}.  
\\

\noindent
{\it Oxygen} -- 
The O$^{2+}$/H$^{+}$ abundances derived from the [O~{\sc iii}] nebular 
lines were adopted for our PNe.  The O$^{2+}$/H$^{+}$ abundance of PN3 
is lower than those of the other PNe.  Given that O$^{2+}$ is one of 
the dominant coolants in PNe, the low O$^{2+}$/H$^{+}$ ionic abundance 
in PN3 makes the cooling inefficient and as a consequence, the highest 
electron temperature of this PN (Table~\ref{temden}) and thus the 
strongest He~{\sc ii} $\lambda$4686 line. 
The [O~{\sc ii}] $\lambda$3727 (a blend of the 
$\lambda\lambda$3726,\,3729 doublet) line is much stronger than the 
[O~{\sc ii}] $\lambda\lambda$7320,\,7330 auroral lines and thus better 
detected.  The O$^{+}$/H$^{+}$ abundance derived from the $\lambda$3727 
line is adopted for all targets. 
\\

\noindent
{\it Nitrogen, Neon} -- 
The weak [N~{\sc ii}] $\lambda$5755 auroral line was detected in most of 
the PNe.  The N$^{+}$/H$^{+}$ abundance derived from the $\lambda$6583 
line was adopted for our targets. 
The [Ne~{\sc iii}] $\lambda$3967 line is blended with H~{\sc i} 
$\lambda$3970, and thus yields higher Ne$^{2+}$/H$^{+}$ abundance than the 
$\lambda$3868 line, which is free of line blending.  The abundance derived 
from [Ne~{\sc iii}] $\lambda$3868 is adopted for all targets. 
\\

\noindent
{\it Argon, Sulfur} -- 
The [Ar~{\sc iv}] $\lambda\lambda$4711,\,4740 doublet is blended with the 
[Ne~{\sc iv}] $\lambda\lambda$4714,15 and 4724,25 lines of the 
2p$^{3}$\,$^{2}$D$^{\rm o}$ -- $^{2}$P$^{\rm o}$ transition.  The flux 
contribution of these [Ne~{\sc iv}] lines are negligible (1--2\%). 
However, the blended He~{\sc i} $\lambda$4713 (2p\,$^{3}$P$^{\rm o}$ -- 
4s\,$^{3}$S) contributes 30--40\% to the total flux at $\lambda$4711 
(Section~\ref{section3:a}).  The corrected flux of the [Ar~{\sc iv}] 
$\lambda$4711 line yielded an Ar$^{3+}$/H$^{+}$ consistent with that 
derived from the $\lambda$4740 line.  The [Ar~{\sc iii}] 
$\lambda\lambda$7136,\,7751 nebular lines were detected in five PNe.  The 
$\lambda$7751 line is located at the red end of the spectra and its flux 
could be unreliable due to low efficiency of the instrument. 
Thus the Ar$^{2+}$/H$^{+}$ abundance derived from the stronger $\lambda$7136 
line is adopted for our targets.  The [Ar~{\sc v}] $\lambda$7005 line was 
detected in PN2, PN3 and PN7, and the Ar$^{4+}$/H$^{+}$ abundance was 
derived using this line (Table~\ref{ionic}).  The S$^{+}$/H$^{+}$ abundance 
derived from the [S~{\sc ii}] $\lambda$6731 line is adopted for our PNe, 
given that $\lambda$6731 is the stronger line of the doublet.  The [S~{\sc 
iii}] $\lambda$6312 line is detected in four of our targets and the 
S$^{2+}$/H$^{+}$ abundance was derived using this line. 
\\

\noindent
{\it Carbon} -- 
The C~{\sc ii} $\lambda$4267 optical recombination line was detected in 
the spectrum of PN2 (Figure~\ref{pn2_carbon}).  Accurate measurements 
of the optical recombination lines of heavy elements were difficult due 
to weakness ($\leq$10$^{-4}$--10$^{-3}$ of the H$\beta$ flux).  In the 
past two decades, deep spectroscopy with an aid of modern high-efficiency 
and large-format linear detectors has enabled detection of many optical 
recombination lines of heavy element ions (mainly C~{\sc ii}, N~{\sc ii}, 
O~{\sc ii}, Ne~{\sc ii}) in the spectra of Galactic PNe, and new 
astrophysics have been revealed through studies of these weak lines 
(e.g., \citealt{liu00}; \citealt{fl13}; see also the reviews 
\citealt{liu06,liu12}). 
The weak optical recombination lines of heavy elements have 
previously been detected in the spectra of extragalactic PNe, e.g., in 
one M31 PNe by \citet[][Table~2]{balick13} and in the Magellanic Cloud 
PNe by \citet{ld06} and \citet{shaw10}. 
The C~{\sc ii} $\lambda$4267 line is produced by radiative recombination 
only, and thus is used to derive the C$^{2+}$/H$^{+}$ abundance in this 
paper.  The C$^{2+}$/H$^{+}$ abundance ratio of target PN2 is 
1.7$\pm$0.8$\times$10$^{-3}$ (Table~\ref{ionic}), which agrees with the 
target observed by \citet{balick13} within errors.

The [O~{\sc iii}] electron temperature of PN2 (11\,600~K) was assumed 
when calculating the recombination line C$^{2+}$/H$^{+}$ abundance.  
This is the same strategy used by \citet{balick13}.  However, deep 
spectroscopic observations of numerous Galactic PNe have provided evidence 
that the optical recombination lines of heavy elements come from plasma 
as cold as $<$1000~K \citep[e.g.,][]{liu06,lbz06,fl13,mcn13}.  If 
this is also true for the M31 PNe, then our C$^{2+}$/H$^{+}$ abundance 
ratio could be overestimated.  At an electron temperature of 1000~K, the 
calculated C$^{2+}$/H$^{+}$ abundance ratio for PN2 is 
1.8$\times$10$^{-4}$, lower than the value calculated using the [O~{\sc 
iii}] temperature by nearly one order of magnitude.  The C~{\sc ii} 
effective recombination coefficients of \citet{dav00} was adopted for 
the abundance calculations. 
\\

The abundance errors following the ionic abundance ratios in 
Table~\ref{ionic} were mainly propagated from the measurement errors 
of the extinction-corrected relative line intensities given in 
Table~\ref{lines}.  Since the [O~{\sc iii}] $\lambda$4363 auroral 
line was well detected in the spectra of all targets 
(Figures~\ref{pn3pn7}), the errors in the [O~{\sc iii}] electron 
temperatures are generally small (Table~\ref{temden}) and not 
considered in the ionic abundance calculations.  However, 
as discussed in Section~\ref{section3:c}, 
adopting the [O~{\sc iii}] temperature for all ionic species might 
introduce extra uncertainties in ionic abundance calculations, 
especially for the low ionization species, as PNe have ionization 
structures.  This uncertainty is difficult to assess due to the 
limited number of reliable temperature diagnostics. 
The errors in the electron temperature potentially usually have 
large impact on in the abundances derived from CELs, given that the 
emissivity of a CEL has an exponential dependence on the electron 
temperature \citep{of06}:  $\epsilon$($\lambda$) $\propto$ 
$T_\mathrm{e}^{-1/2}$\,$\exp$($-E_{\rm ex}$/$kT_\mathrm{e}$), where 
$E_{\rm ex}$ is the excitation energy of the upper level of a CEL 
transition.  We estimated that the errors in our [O~{\sc iii}] 
temperatures typically introduce $\sim$9--12\% uncertainties in the 
CEL ionic abundances.

\begin{table*}
\rotatebox{90}{
\begin{minipage}{\textheight}
\begin{center}
\caption{Ionic Abundances}
\label{ionic}
\begin{tabular}{llccccccc}
\hline
\hline
Ion & Line  & 
\multicolumn{7}{c}{\underline{~~~~~~~~~~~~~~~~~~~~~~~~~~~~~~~~~~~~~~~~~~~~~~~~~~~~~~~~~~~~~~~~~~~~~~~~~~~~~~ 
Abundance (X$^{i+}$/H$^{+}$) ~~~~~~~~~~~~~~~~~~~~~~~~~~~~~~~~~~~~~~~~~~~~~~~~~~~~~~~~~~~~~~~~~~~~~~~~~~~~~~}}
\\
    & (\AA) & PN1  &  PN2  &  PN3  &  PN4  &  PN5  &  PN6  &  PN7\\
\hline
He$^{+}$ &  3888 &    0.059$\pm$0.027 &    0.057$\pm$0.026 &    0.062$\pm$0.018 &    0.049$\pm$0.023 &    0.053$\pm$0.024 &    0.082$\pm$0.031 &    0.076$\pm$0.025\\
         &  4471 &    0.112$\pm$0.018 &    0.124$\pm$0.020 &    0.102$\pm$0.012 &    0.124$\pm$0.020 &    0.168$\pm$0.034 &    0.097$\pm$0.024 &    0.131$\pm$0.014\\
         &  5876 &    0.113$\pm$0.012 &    0.123$\pm$0.014 &    0.097$\pm$0.010 &    0.101$\pm$0.011 &    0.112$\pm$0.015 &    0.087$\pm$0.015 &    0.106$\pm$0.006\\
         &  6678 &    0.122$\pm$0.019 &    0.145$\pm$0.023 &    0.099$\pm$0.014 &    0.112$\pm$0.017 &    0.096$\pm$0.013 &    ...             &    0.109$\pm$0.016\\
\multicolumn{2}{l}{Adopted$^{a}$} &  0.113$\pm$0.012 &  0.123$\pm$0.014 &  0.097$\pm$0.010 &  0.101$\pm$0.011 &  0.112$\pm$0.015 &  0.087$\pm$0.015 &  0.106$\pm$0.006\\
He$^{2+}$ & 4686 & 0.63($\pm$0.26)$\times$10$^{-3}$ & 3.40($\pm$0.49)$\times$10$^{-3}$ & 1.58($\pm$0.14)$\times$10$^{-2}$ & 1.17($\pm$0.36)$\times$10$^{-3}$ & 1.63($\pm$0.39)$\times$10$^{-3}$ & 1.53($\pm$:)$\times$10$^{-3}$ & 1.89($\pm$0.28)$\times$10$^{-3}$\\
C$^{2+}$  & 4267 &  ...                             & 1.7($\pm$0.8)$\times$10$^{-3}$   & ...                              & ...                              & ...                              & ...                           & ...                             \\
N$^{+}$  & 5755 & 1.46($\pm$0.60)$\times$10$^{-5}$ & 1.27($\pm$0.39)$\times$10$^{-5}$ & 9.23($\pm$2.10)$\times$10$^{-6}$ & 9.12($\pm$2.45)$\times$10$^{-6}$ & 1.76($\pm$0.74)$\times$10$^{-5}$ & ...                           & 1.08($\pm$0.32)$\times$10$^{-5}$\\
         & 6548 & 8.51($\pm$1.02)$\times$10$^{-6}$ & 1.34($\pm$0.20)$\times$10$^{-5}$ & 1.01($\pm$0.13)$\times$10$^{-5}$ & 7.84($\pm$1.28)$\times$10$^{-6}$ & 1.39($\pm$0.34)$\times$10$^{-5}$ & 2.95($\pm$1.32)$\times$10$^{-6}$ & 1.39($\pm$0.17)$\times$10$^{-5}$\\
         & 6583 & 8.91($\pm$0.97)$\times$10$^{-6}$ & 1.35($\pm$0.21)$\times$10$^{-5}$ & 1.01($\pm$0.15)$\times$10$^{-5}$ & 8.02($\pm$1.30)$\times$10$^{-6}$ & 1.37($\pm$0.35)$\times$10$^{-5}$ & 3.03($\pm$1.30)$\times$10$^{-6}$ & 1.47($\pm$0.20)$\times$10$^{-5}$\\
\multicolumn{2}{l}{Adopted$^{b}$} &  8.91($\pm$0.97)$\times$10$^{-6}$ & 1.35($\pm$0.21)$\times$10$^{-5}$ & 1.01($\pm$0.15)$\times$10$^{-5}$ & 8.02($\pm$1.30)$\times$10$^{-6}$ & 1.37($\pm$0.35)$\times$10$^{-5}$ & 3.03($\pm$1.30)$\times$10$^{-6}$ & 1.47($\pm$0.20)$\times$10$^{-5}$\\
O$^{+}$  & 3727 & 2.05($\pm$0.36)$\times$10$^{-5}$ & 2.52($\pm$0.61)$\times$10$^{-5}$ & 1.98($\pm$0.40)$\times$10$^{-5}$ & 2.06($\pm$0.49)$\times$10$^{-5}$ & 2.35($\pm$0.66)$\times$10$^{-5}$ & 2.96($\pm$0.78)$\times$10$^{-5}$ & 5.1($\pm$0.9)$\times$10$^{-5}$\\
         & 7320 & 6.13($\pm$1.54)$\times$10$^{-5}$ & 6.17($\pm$1.65)$\times$10$^{-5}$ & 2.54($\pm$0.70)$\times$10$^{-5}$ & 3.02($\pm$0.96)$\times$10$^{-5}$ & ...                              & ...                              & 5.1($\pm$1.2)$\times$10$^{-5}$\\
         & 7330 & 7.24($\pm$1.82)$\times$10$^{-5}$ & 8.50($\pm$2.27)$\times$10$^{-5}$ & 2.14($\pm$0.58)$\times$10$^{-5}$ & 2.78($\pm$0.91)$\times$10$^{-5}$ & ...                              & ...                              & 6.42($\pm$1.51)$\times$10$^{-5}$\\
\multicolumn{2}{l}{Adopted$^{c}$} &  2.05($\pm$0.36)$\times$10$^{-5}$ & 2.52($\pm$0.61)$\times$10$^{-5}$ & 1.98($\pm$0.40)$\times$10$^{-5}$ & 2.06($\pm$0.49)$\times$10$^{-5}$ & 2.35($\pm$0.66)$\times$10$^{-5}$ & 2.96($\pm$0.78)$\times$10$^{-5}$ & 5.1($\pm$0.9)$\times$10$^{-5}$\\
O$^{2+}$ & 4363 & 2.58($\pm$0.60)$\times$10$^{-4}$ & 2.73($\pm$0.57)$\times$10$^{-4}$ & 2.19($\pm$0.43)$\times$10$^{-4}$ & 2.02($\pm$0.48)$\times$10$^{-4}$ & 3.61($\pm$1.05)$\times$10$^{-4}$ & 4.45($\pm$1.52)$\times$10$^{-4}$ & 3.17($\pm$0.64)$\times$10$^{-4}$\\
         & 4959 & 2.50($\pm$0.24)$\times$10$^{-4}$ & 2.68($\pm$0.31)$\times$10$^{-4}$ & 2.18($\pm$0.24)$\times$10$^{-4}$ & 1.99($\pm$0.32)$\times$10$^{-4}$ & 3.44($\pm$0.56)$\times$10$^{-4}$ & 4.47($\pm$0.84)$\times$10$^{-4}$ & 3.08($\pm$0.33)$\times$10$^{-4}$\\
         & 5007 & 2.60($\pm$0.26)$\times$10$^{-4}$ & 2.75($\pm$0.32)$\times$10$^{-4}$ & 2.20($\pm$0.25)$\times$10$^{-4}$ & 2.03($\pm$0.37)$\times$10$^{-4}$ & 3.51($\pm$0.57)$\times$10$^{-4}$ & 4.49($\pm$0.86)$\times$10$^{-4}$ & 3.20($\pm$0.34)$\times$10$^{-4}$\\
\multicolumn{2}{l}{Adopted$^{d}$} &  2.60($\pm$0.26)$\times$10$^{-4}$ & 2.75($\pm$0.32)$\times$10$^{-4}$ & 2.20($\pm$0.25)$\times$10$^{-4}$ & 2.03($\pm$0.37)$\times$10$^{-4}$ & 3.57($\pm$0.57)$\times$10$^{-4}$ & 4.49($\pm$0.86)$\times$10$^{-4}$ & 3.20($\pm$0.34)$\times$10$^{-4}$\\
Ne$^{2+}$ & 3868 & 5.85($\pm$0.66)$\times$10$^{-5}$ & 7.31($\pm$0.97)$\times$10$^{-5}$ & 4.62($\pm$0.59)$\times$10$^{-5}$ & 4.69($\pm$0.76)$\times$10$^{-5}$ & 9.48($\pm$1.78)$\times$10$^{-5}$ & 1.07($\pm$0.23)$\times$10$^{-4}$ & 7.61($\pm$0.96)$\times$10$^{-5}$\\
          & 3967 & 7.06($\pm$1.17)$\times$10$^{-5}$ & 8.22($\pm$1.40)$\times$10$^{-5}$ & 5.82($\pm$1.00)$\times$10$^{-5}$ & 5.46($\pm$0.88)$\times$10$^{-5}$ & 1.09($\pm$0.20)$\times$10$^{-4}$ & 1.12($\pm$0.25)$\times$10$^{-4}$ & 9.36($\pm$1.22)$\times$10$^{-5}$\\
\multicolumn{2}{l}{Adopted$^{e}$} & 5.85($\pm$0.66)$\times$10$^{-5}$ & 7.31($\pm$0.97)$\times$10$^{-5}$ & 4.62($\pm$0.59)$\times$10$^{-5}$ & 4.69($\pm$0.76)$\times$10$^{-5}$ & 9.54($\pm$1.78)$\times$10$^{-5}$ & 1.07($\pm$0.23)$\times$10$^{-4}$ & 7.61($\pm$0.96)$\times$10$^{-5}$\\
S$^{+}$   & 6716 & 1.34($\pm$0.28)$\times$10$^{-7}$ & 2.37($\pm$0.45)$\times$10$^{-7}$ & 3.30($\pm$0.36)$\times$10$^{-7}$ & 3.77($\pm$0.64)$\times$10$^{-7}$ & 4.84($\pm$0.74)$\times$10$^{-7}$ & ...                              & 1.01($\pm$0.11)$\times$10$^{-6}$\\
          & 6731 & 1.60($\pm$0.30)$\times$10$^{-7}$ & 3.08($\pm$0.52)$\times$10$^{-7}$ & 3.44($\pm$0.38)$\times$10$^{-7}$ & 3.43($\pm$0.60)$\times$10$^{-7}$ & 4.61($\pm$0.67)$\times$10$^{-7}$ & ...                              & 8.75($\pm$1.05)$\times$10$^{-7}$\\
\multicolumn{2}{l}{Adopted$^{f}$} & 1.60($\pm$0.30)$\times$10$^{-7}$ & 3.08($\pm$0.52)$\times$10$^{-7}$ & 3.44($\pm$0.38)$\times$10$^{-7}$ & 3.43($\pm$0.60)$\times$10$^{-7}$ & 4.61($\pm$0.67)$\times$10$^{-7}$ & ...                              & 8.75($\pm$1.05)$\times$10$^{-7}$\\
S$^{2+}$  & 6312 & 1.92($\pm$0.33)$\times$10$^{-6}$ & 2.93($\pm$0.58)$\times$10$^{-6}$ & 1.09($\pm$0.18)$\times$10$^{-6}$ & ...                              & ...                              & ...                              & 3.7($\pm$0.7)$\times$10$^{-6}$\\
Ar$^{2+}$ & 7136 & 1.14($\pm$0.20)$\times$10$^{-6}$ & 1.06($\pm$0.23)$\times$10$^{-6}$ & 1.12($\pm$0.21)$\times$10$^{-6}$ & 1.01($\pm$0.28)$\times$10$^{-6}$ & 1.89($\pm$0.47)$\times$10$^{-6}$ & ...                              & 1.5($\pm$0.4)$\times$10$^{-6}$\\
          & 7751 & 9.65($\pm$3.38)$\times$10$^{-7}$ & 3.72($\pm$1.41)$\times$10$^{-7}$ & 1.33($\pm$0.52)$\times$10$^{-6}$ & 1.13($\pm$0.51)$\times$10$^{-6}$ & 2.33($\pm$0.88)$\times$10$^{-6}$ & ...                              & 1.3($\pm$0.7)$\times$10$^{-6}$\\
\multicolumn{2}{l}{Adopted$^{g}$} & 1.14($\pm$0.20)$\times$10$^{-6}$ & 1.06($\pm$0.23)$\times$10$^{-6}$ & 1.12($\pm$0.21)$\times$10$^{-6}$ & 1.01($\pm$0.28)$\times$10$^{-6}$ & 1.89($\pm$0.47)$\times$10$^{-6}$ & ...                              & 1.51($\pm$0.36)$\times$10$^{-6}$\\
Ar$^{3+}$ & 4711 & 1.52($\pm$0.35)$\times$10$^{-7}$ & 1.36($\pm$0.31)$\times$10$^{-7}$ & 4.67($\pm$0.80)$\times$10$^{-7}$ & 2.57($\pm$0.67)$\times$10$^{-7}$ & 4.83($\pm$1.35)$\times$10$^{-7}$ & ...                              & 2.8($\pm$0.6)$\times$10$^{-7}$\\
          & 4740 & 1.52($\pm$0.27)$\times$10$^{-7}$ & 1.34($\pm$0.30)$\times$10$^{-7}$ & 4.68($\pm$0.75)$\times$10$^{-7}$ & 2.58($\pm$0.65)$\times$10$^{-7}$ & 4.84($\pm$1.30)$\times$10$^{-7}$ & ...                              & 2.8($\pm$0.5)$\times$10$^{-7}$\\
\multicolumn{2}{l}{Adopted$^{h}$} & 1.52($\pm$0.30)$\times$10$^{-7}$ & 1.35($\pm$0.30)$\times$10$^{-7}$ & 4.68($\pm$0.77)$\times$10$^{-7}$ & 2.58($\pm$0.66)$\times$10$^{-7}$ & 4.84($\pm$1.33)$\times$10$^{-7}$ &  ...                             & 2.8($\pm$0.5)$\times$10$^{-7}$\\
Ar$^{4+}$ & 7005 & ...                              & 8.00($\pm$2.83)$\times$10$^{-8}$ & 6.82($\pm$2.25)$\times$10$^{-8}$ & ...                              & ...                              & ...                              & 7.6($\pm$2.3)$\times$10$^{-8}$\\
\hline
\end{tabular}
\end{center}
\begin{description}
\item[$^{a}$] The He$^{+}$/H$^{+}$ abundance derived from the He~{\sc i}
$\lambda$5876 line is adopted for all PNe.
\item[$^{b}$] The N$^{+}$/H$^{+}$ abundance derived from the [N~{\sc ii}]
$\lambda$6583 line is adopted for all PNe.
\item[$^{c}$] 
The O$^{+}$/H$^{+}$ abundance derived from the [O~{\sc ii}] $\lambda$3727 
doublet is adopted for all PNe. 
\item[$^{d}$] The O$^{2+}$/H$^{+}$ abundance derived from the [O~{\sc iii}]
$\lambda$5007 line was adopted for all PNe.
\item[$^{e}$] The Ne$^{2+}$/H$^{+}$ abundance derived from the [Ne~{\sc
iii}] $\lambda$3868 line is adopted.  [Ne~{\sc iii}] $\lambda$3967 is
blended with H~{\sc i} $\lambda$3970.
\item[$^{f}$] The S$^{+}$/H$^{+}$ abundance derived from the [S~{\sc ii}]
$\lambda$6731 line is adopted, because the critical density of $\lambda$6731
is much higher than that of the $\lambda$6716 line.
\item[$^{g}$] The Ar$^{2+}$/H$^{+}$ abundance derived from the [Ar~{\sc
iii}] $\lambda$7136 line is adopted.  Measurements of $\lambda$7136 are
more reliable than those of the $\lambda$7751 line at the red end of the
spectra.
\item[$^{h}$] The Ar$^{3+}$/H$^{+}$ abundance derived from the total
intensity of [Ar~{\sc iv}] $\lambda\lambda$4711,4740 is adopted for all
PNe.
\end{description}
\end{minipage}}
\end{table*}

\subsubsection{Elemental Abundances} \label{section3:d:2}

The total elemental abundances derived for the seven PNe are presented 
in Table~\ref{elemental}, together with the Solar abundances from 
\citet{asp09}.  The He/H elemental abundances of our targets were 
calculated by adding the He$^{+}$/H$^{+}$ and He$^{2+}$/H$^{+}$ ratios. 
Whenever available, the ionization correction factors (ICFs) given by 
\citet{kb94} were used to calculate the total abundances of the heavy 
elements.  The ICF values used to calculate the elemental abundances 
for the seven PNe are given in Table~\ref{icfs}. 
It is worth mentioning that when calculating the elemental abundance 
of argon, depending on the detection or absence of the [Ar~{\sc v}] 
$\lambda$7005 line, different ICFs have been used for our targets 
(Equations~A30 and A34 in \citealt{kb94}).  The same scheme was adopted 
in the calculations of sulfur:  The ICFs used for the targets, where the 
[S~{\sc iii}] $\lambda$6312 auroral line was detected, are different 
from those where the [S~{\sc iii}] line was not detected (Equations~15 
and A36 in \citealt{kb94}).

Errors in the brackets following the total elemental abundances in 
Table~\ref{elemental} were estimated based on the errors in the ionic 
abundances through a simple propagation paradigm.  For helium, the 
error is mainly contributed by uncertainties in the He$^{+}$/H$^{+}$ 
ratio, whose concentration is much higher than that of He$^{2+}$ 
(Table~\ref{ionic}).  For heavy elements, the errors can also be 
introduced by the use of ICFs.  This source of error is negligible for 
oxygen, whose ICF is always close to unity (Table~\ref{icfs}).  For 
other heavy elements, uncertainties introduced by ICFs could be 
significant.  Abundances of nitrogen and neon were derived based on the 
ionic and elemental abundances of oxygen, and thus are expected to be 
reliable.  The total sulfur abundance is generally quite uncertain, as 
only the [S~{\sc ii}] lines are well observed for this element. 
Although the [S~{\sc iii}] $\lambda$6312 line was also detected in the 
spectra of some PNe, it arises from an auroral transition 
(3p$^{2}$~$^{1}$D$_{2}$ -- 3p$^{2}$~$^{1}$S$_{0}$), hence it is 
particularly temperature sensitive.  Measurements of the [S~{\sc iii}] 
$\lambda$6312 line is affected by the subtraction of the bright [O~{\sc 
i}] $\lambda$6300 skyline.  Besides, this [S~{\sc iii}] line is also 
blended with the He~{\sc ii} $\lambda$6311 (5g~$^{2}$G -- 
16h~$^{2}$H$^{\rm o}$) line.  Uncertainties and systematic errors in 
the ICFs are difficult to define, and thus they were not considered for 
the error estimate of our objects.  The actual uncertainties in the 
elemental abundances of nitrogen, neon, sulfur and argon in 
Table~\ref{elemental} must be regarded as lower limits of the real 
abundance uncertainties.

\subsection{Duplication with Recent Spectroscopic Surveys} \label{section3:e}

Of all our targets, including the three PNe studied in Paper~I, only PN4 
(M77) was observed by \citet{san12}, who obtained spectra for 253 H~{\sc 
ii} regions and 407 PNe in M31 using the Hectospec multi-fiber spectrograph 
attached to the 6.5\,m MMT.  Oxygen abundance was derived for 51 PNe in the 
sample of \citet{san12} based on the electron temperatures estimated from 
the [O~{\sc iii}] $\lambda$4363 auroral line.  However, the $\lambda$4363 
line was not detected in the MMT spectrum of M77, and thus the O/H ratio 
was not estimated.  For the most prominent nebular emission lines, fluxes 
measured in our GTC spectrum of M77 only differ slightly from those 
of the MMT spectrum:  2\% for the [O~{\sc iii}] $\lambda\lambda$4959,\,5007 
lines, 7\% for the [N~{\sc ii}] $\lambda\lambda$6548,\,6583 lines, 12\% 
for the [O~{\sc ii}] $\lambda$3727 lines, and 20--40\% for the [S~{\sc ii}] 
$\lambda\lambda$6716,\,6731 lines.




\section{Discussion} \label{section4}

\subsection{The Spatial and Kinematical Distribution} \label{section4:a}

\begin{figure*}
\begin{center}
\includegraphics[width=16.0cm,angle=0]{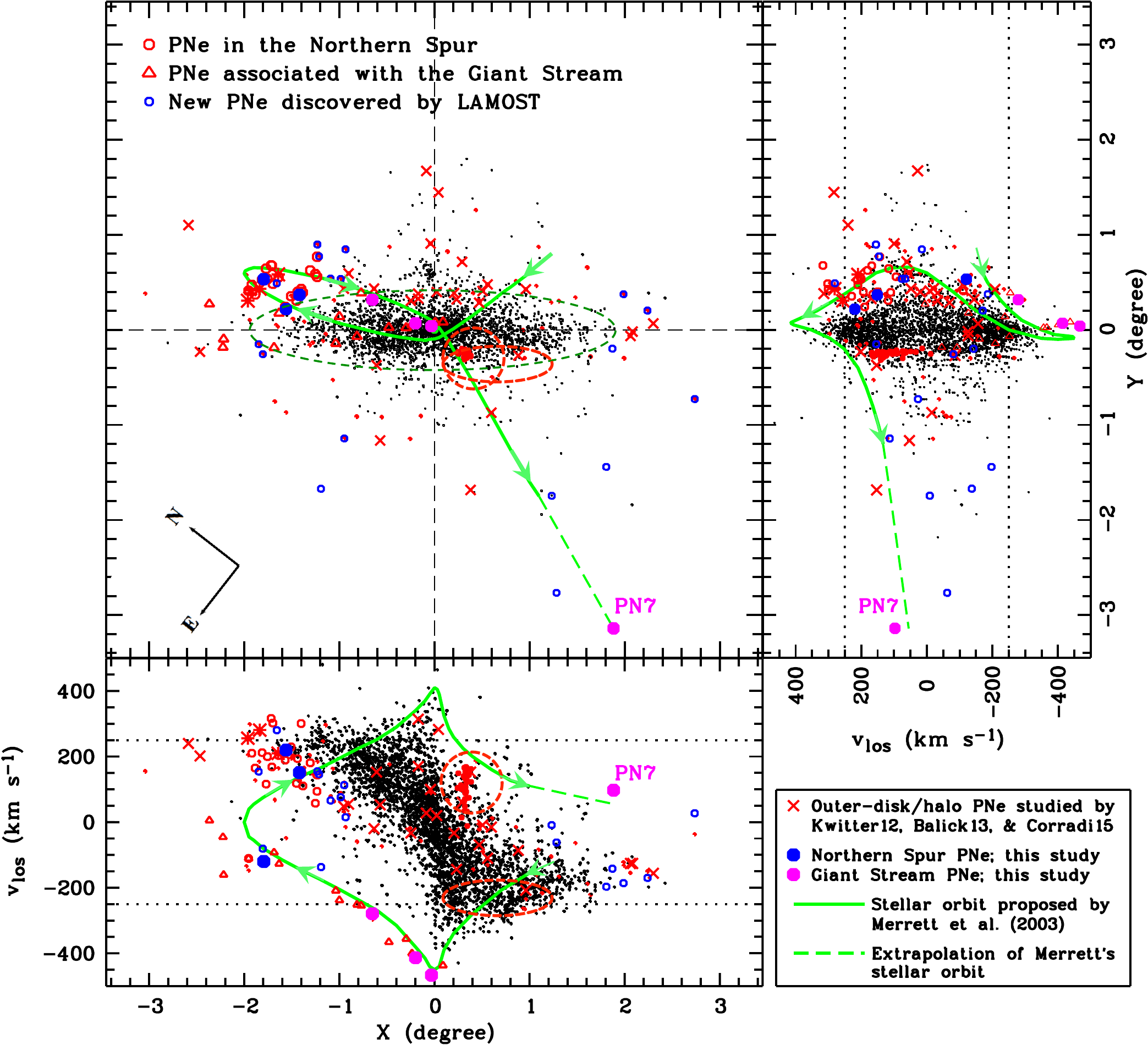}
\caption{Spatial distribution and kinematics of PNe in M31 (see the 
description of the $X$ and $Y$ coordinates in the text).  PNe samples are 
from \citet[][black dots; including the sample of \citealt{hkpn04}]{mer06}, 
\citet[][red dots]{kniazev14}, and the LAMOST survey \citep[][blue open 
circles]{yuan10}.  Red open circles represent the Northern Spur PNe and 
red open triangles are those associated with the Giant Stream.  The three 
Northern Spur PNe studied in Paper~I are the red asterisks.  The seven 
PNe in this study are highlighted by the blue and magenta filled circles 
(see the bottom-right legend, also Figure~\ref{m31_pne}), and PN7, a 
distant PN newly identified by LAMOST (see Section~\ref{section2:a}), is 
marked.  Red crosses are the outer-disk and halo PNe observed by 
\citet{kwi12}, \citet{balick13}, and \citet{cor15}.  In the upper left 
panel: overplotted is the stellar orbit (thick green curve) proposed by 
\citet{mer03}; the green dashed ellipse has a semimajor axis of 2$^{\circ}$ 
($\sim$27~kpc) and represents a disk with an inclination angle $i$ = 
77\fdg7 \citep{dev58}.  In the two side panels (bottom and right):  a 
projection of the orbit in line-of-sight velocity with respect to M31, 
$v_{\rm los}$, vs. distance along the major and minor axes of M31 is 
superimposed on the PNe data.  All velocities of PNe are corrected for 
the systemic velocity $-$306 km\,s$^{-1}$ \citep{corbelli10}.  Arrows on 
the orbit show the direction along the stream.  The red dashed circle and 
ellipse show the locations of satellites M32 and And~{\sc viii}, 
respectively, and the red dots in the red dashed circle are the PNe 
belonging to M32.  The thick green dashed lines are tangential 
extrapolations of the stellar orbit. 
The black dotted lines at $\pm$250~km\,s$^{-1}$ \citep{corbelli10} 
show the adopted circular speed of M31 disk. 
The format of the figure generally follows Figure\,2 of \citet[][also 
Figure~17 in Paper~I]{mer03} and the stellar orbit was reproduced based 
on that figure (with permission of the authors). }
\label{orbit}
\end{center}
\end{figure*}

Modern large-area surveys have revealed that M31 has an extended stellar 
disk and a huge halo \citep[$\geq$300~kpc;][]{iba05,iba07,iba14}.  The 
outer regions of M31 are extremely complex.  Numerous structures such as 
streams, loops, and overdensity regions were discovered throughout the halo 
\citep[e.g.,][]{mccon09,lewis13,iba14}.  These features, with the Northern 
Spur and the Giant Stream being the most prominent, are mostly associated 
with accretion/interaction of satellite dwarf galaxies. 
Although the possible connection between the Northern Spur and the Giant 
Stream had already been inferred by \citet{fer02} and \citet{mccon03}, 
\citet{mer03} were the first to explicitly propose this connection. 
Based on a kinematic study of $\sim$20 PNe in the disk of M31, 
\citet{mer03} presented a possible orbit for the stellar stream in M31, 
which connects the Giant Stream to the Northern Spur.  The Northern Spur 
is located at the turning point of this model orbit, which is strongly 
warped as the stellar stream passes nearby the center of M31 
(Figure~\ref{orbit}, upper-left panel). 

Figure~\ref{orbit} shows the positions of PNe observed by \citet{mer06} 
in the $X$--$Y$ coordinate system in an M31-based reference frame, 
where $X$ lies along the major axis of M31 and increases toward the 
southwest, and $Y$ lies along the minor axis and increases toward the 
northwest.  Both coordinates were calculated following the geometric 
transformations of \citet{hbk91}.  The orbit proposed by \citet{mer03} 
is presented along with the PNe in Figure~\ref{orbit}.  The two side 
(bottom and right) panels in Figure~\ref{orbit} shows the projection 
of this stellar orbit in the line-of-sight velocity with respect to M31 
($v_{\rm los}$) versus distance along the major and minor axes.  PNe in 
the Northern Spur and those associated with the Giant Stream are 
highlighted.  The seven PNe in this study and the three observed in 
Paper~I are also highlighted (see explanation of symbols in the caption 
of Figure~\ref{orbit}).

The PNe in the two substructures are generally associated with the 
stellar orbit of \citet{mer03}, and this association is particularly 
consistent for those located on the Giant Stream.  The three Northern 
Spur PNe observed in Paper~I (see the discussion therein) are located 
at the turning point of the orbit.  The positions of the seven objects 
targeted by the current work show excellent agreement with the orbit, 
both spatially and kinematically.  The six PNe, named PN1--PN6, as 
carefully selected from the catalog of \citet{mer03}, were identified 
to be located in the two substructures by \citet{mer06}.  It is worth 
noting that the spatial extent of the stellar orbit was originally 
confined to within $\sim$4$^{\circ}$\,$\times$\,4$^{\circ}$ centered 
on M31 (see Figure~2 of \citealt{mer03}).  The newly discovered PN7, 
with $v_{\rm los}$ measured by the LAMOST spectroscopy, was located 
3\fdg65 from M31's center, beyond the orbit's extent of \citet{mer03}. 
However, the spatial position of PN7 matches well the extrapolation of 
the projected orbit in the tangential direction (the thick green dashed 
line in Figure~\ref{orbit}).

Although slight deviation between PN7 and the extrapolated orbit exists 
in the $v_{\rm los}$ vs. $X$ and $v_{\rm los}$ vs. $Y$ diagrams 
(Figure~\ref{orbit}, bottom and right panels), the agreement is still 
reasonable within the possible dispersion of the orbit, as \citet{mer03} 
claimed that the stellar stream is probably not in reality a single 
orbit, but a family of adjacent orbits, and the singular nature of the 
potential of their simplified model also amplifies the dispersion in 
the orbit and increases the spread in the observed $v_{\rm los}$ to 
$\sim$100~km\,s$^{-1}$.  This model orbit is only a generic 
representative of the stellar streams that connect the two photometric 
features.

The kinematics of the 27 outer-disk/halo PNe observed by 
\citet[][Kwitter12; 16 PNe]{kwi12}, \citet[][Balick13; 2 PNe]{balick13}, 
and \citet[][Corradi15; 9 PNe]{cor15} generally resemble the rotation 
pattern of the classical disk of M31 (Figure~\ref{orbit}, bottom and 
right panels), and can be kinematically distinguished from our sample, 
which belong to the substructures and are mostly well located on the 
stellar stream of \citet{mer03}.  Even for the PNe at very large radii, 
e.g., those along the major axis of M31 observed by \citet{cor15}, the 
kinematics still follow the extended disk.  The kinematics of PNe in 
the Northern Spur (shown as open circles in Figure~\ref{orbit}) are 
indistinguishable from those of the disk, although this substructure is 
located some distance from the plane. 

\begin{table*}
\begin{center}
\caption{Elemental Abundances$^{a}$}
\label{elemental}
\begin{tabular}{lcc|cc|cc|cc}
\hline
\hline
Elem. & \multicolumn{8}{c}{\underline{~~~~~~~~~~~~~~~~~~~~~~~~~~~~~~~~~~~~~~~~~~~~~~~~~~~~~~~~~~~~~~~~~~~
X/H ~~~~~~~~~~~~~~~~~~~~~~~~~~~~~~~~~~~~~~~~~~~~~~~~~~~~~~~~~~~~~~~~~~~}}\\
        & \multicolumn{2}{c}{~~~~PN1~~~~} & \multicolumn{2}{c}{~~~~PN2~~~~} & \multicolumn{2}{c}{~~~~PN3~~~~} & \multicolumn{2}{c}{~~~~PN4~~~~}\\
\hline
He & 0.113$\pm$0.013                  & 11.05 & 0.126$\pm$0.017                  & 11.10 & 0.113$\pm$0.014                  & 11.05 & 0.102$\pm$0.012                  & 11.01\\
C  & ...                              & ...   & 2.15($\pm$0.73)$\times$10$^{-3}$ &  9.33 & ...                              & ...   & ...                              & ...  \\
N  & 4.39($\pm$1.10)$\times$10$^{-5}$ &  7.64 & 6.60($\pm$2.77)$\times$10$^{-5}$ &  7.82 & 1.15($\pm$0.50)$\times$10$^{-4}$ &  8.06 & 6.44($\pm$2.83)$\times$10$^{-5}$ &  7.81\\
O  & 2.81($\pm$0.93)$\times$10$^{-4}$ &  8.45 & 3.06($\pm$0.95)$\times$10$^{-4}$ &  8.49 & 2.69($\pm$0.57)$\times$10$^{-4}$ &  8.43 & 2.34($\pm$0.77)$\times$10$^{-4}$ &  8.37\\
Ne & 7.36($\pm$1.40)$\times$10$^{-5}$ &  7.87 & 9.39($\pm$3.94)$\times$10$^{-5}$ &  7.97 & 5.65($\pm$2.54)$\times$10$^{-5}$ &  7.75 & 5.40($\pm$2.38)$\times$10$^{-5}$ &  7.73\\
S  & 1.52($\pm$0.68)$\times$10$^{-6}$ &  6.18 & 2.90($\pm$1.30)$\times$10$^{-6}$ &  6.46 & 4.60($\pm$2.53)$\times$10$^{-6}$ &  6.66 & 3.98($\pm$2.23)$\times$10$^{-6}$ &  6.60\\
Ar & 1.62($\pm$0.73)$\times$10$^{-6}$ &  6.21 & 1.60($\pm$0.75)$\times$10$^{-6}$ &  6.20 & 1.82($\pm$0.80)$\times$10$^{-6}$ &  6.26 & 1.44($\pm$0.76)$\times$10$^{-6}$ &  6.16\\
\hline
Elem. & \multicolumn{8}{c}{\underline{~~~~~~~~~~~~~~~~~~~~~~~~~~~~~~~~~~~~~~~~~~~~~~~~~~~~~~~~~~~~~~~~~~~
X/H ~~~~~~~~~~~~~~~~~~~~~~~~~~~~~~~~~~~~~~~~~~~~~~~~~~~~~~~~~~~~~~~~~~~}}\\
        & \multicolumn{2}{c}{~~~~PN5~~~~} & \multicolumn{2}{c}{~~~~PN6~~~~} & \multicolumn{2}{c}{~~~~PN7~~~~} & \multicolumn{2}{c}{Solar$^{b}$}\\
\hline
He & 0.114$\pm$0.015                  & 11.05 & 0.089$\pm$0.015                  & 10.95 & 0.108$\pm$0.012                  & 11.03 & 0.085                 & 10.93\\
C  & ...                              & ...   & ...                              & ...   & ...                              & ...   & 2.69$\times$10$^{-4}$ &  8.43\\
N  & 1.13($\pm$0.43)$\times$10$^{-4}$ &  8.05 & 4.96($\pm$1.73)$\times$10$^{-5}$ &  7.70 & 9.84($\pm$2.75)$\times$10$^{-5}$ &  7.99 & 6.76$\times$10$^{-5}$ &  7.83\\
O  & 4.11($\pm$1.20)$\times$10$^{-4}$ &  8.61 & 4.84($\pm$1.31)$\times$10$^{-4}$ &  8.68 & 3.81($\pm$0.96)$\times$10$^{-4}$ &  8.58 & 4.89$\times$10$^{-4}$ &  8.69\\
Ne & 1.10($\pm$0.48)$\times$10$^{-4}$ &  8.04 & 1.16($\pm$0.48)$\times$10$^{-4}$ &  8.06 & 9.07($\pm$3.53)$\times$10$^{-5}$ &  7.96 & 8.51$\times$10$^{-5}$ &  7.93\\
S  & 1.07($\pm$0.61)$\times$10$^{-5}$ &  7.03 & ...                              &  ...  & 9.38($\pm$4.13)$\times$10$^{-6}$ &  6.97 & 1.32$\times$10$^{-5}$ &  7.12\\
Ar & 2.71($\pm$1.30)$\times$10$^{-6}$ &  6.43 & ...                              &  ...  & 2.20($\pm$0.83)$\times$10$^{-6}$ &  6.34 & 2.51$\times$10$^{-6}$ &  6.40\\
\hline
\end{tabular}
\begin{description}
\item[$^{a}$] For each PN, the abundances in the left column are linear 
and those in the right column are logarithm, $\log$(X/H) + 12.
\item[$^{b}$] \citet{asp09}.
\end{description}
\end{center}
\end{table*}

\begin{table}
\begin{center}
\begin{minipage}{90mm}
\caption{Ionization Correction Factors}
\label{icfs}
\begin{tabular}{lrrrrrrr}
\hline
\hline
Elem. & \multicolumn{7}{c}{\underline{~~~~~~~~~~~~~~~~~~~~~~~~~~~~~~~~
ICF ~~~~~~~~~~~~~~~~~~~~~~~~~~~~~~~~}}\\
      & PN1 & PN2 & PN3 & PN4 & PN5 & PN6 & PN7\\
\hline
He &  1.000 &  1.000 &  1.000 &  1.000 &  1.000 &  1.000 &  1.000\\
C  &  ...   &  1.285 &  ...   &  ...   &  ...   &  ...   &  ...  \\
N  &  4.932 &  4.900 & 11.412 &  8.029 &  8.218 & 16.354 &  6.705\\
O  &  1.002 &  1.018 &  1.105 &  1.008 &  1.009 &  1.011 &  1.011\\
Ne &  1.258 &  1.285 &  1.224 &  1.152 &  1.151 &  1.078 &  1.191\\
S  &  1.266 &  1.263 &  1.608 &  1.448 &  1.458 &  1.796 &  1.375\\
Ar &  1.254 &  1.256 &  1.096 &  1.142 &  1.138 &  1.065 &  1.175\\
\hline
\end{tabular}
\end{minipage}
\end{center}
\end{table}

\subsection{Abundance Correlations} \label{section4:b}

PNe probes the past chemical composition of the interstellar medium 
(ISM).  Abundances of the $\alpha$-elements such as O, Ne, S, and Ar, 
of a PN represent the chemistry of the ISM in the era when the progenitor 
star just formed.  On the other hand, H~{\sc ii} regions trace the 
current abundances of the ISM.  Giant extragalactic H~{\sc ii} regions, 
where massive star formation activities occur, are amongst the most 
prominent features seen in a gas-rich, star-forming galaxy.  They 
provide probes of relatively homogeneous interstellar material which 
generally show little evidence of abundance enhancements produced by 
the evolved stars embedded in them \citep[e.g.,][]{wof09}.  A comparison 
of the abundances of PNe and H~{\sc ii} regions at the same site (e.g., 
on the galaxy disk) is useful for studying the enrichment history of 
galaxies \citep[e.g.,][]{stan10}. 
Investigation of relations between the abundances of different 
$\alpha$-elements helps to constrain the stellar evolution models 
and quantify the relative yields of each element in the asymptotic giant 
branch (AGB) stars, the immediate progenitors of PNe 
\citep[e.g.,][]{kwi01,mil02,kwi03,hen04}. 
The latest review on the status of modeling the evolution and 
nucleosynthesis of AGB stars was presented by \citet{herwig05}.

\begin{figure*}
\begin{center}
\includegraphics[width=1.95\columnwidth,angle=0]{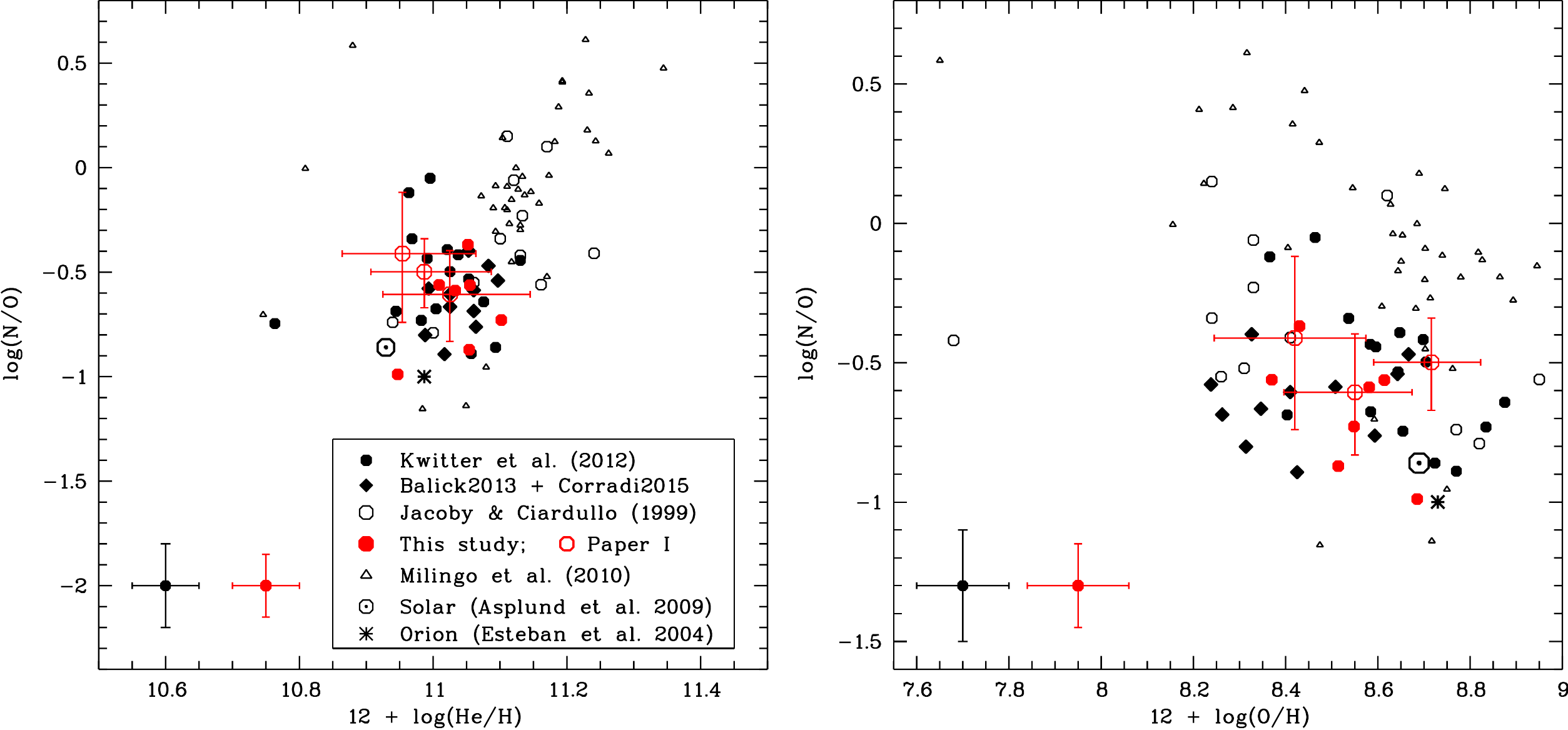}
\caption{Left: $\log$(N/O) vs.\ 12\,+\,$\log$(He/H). 
Right: $\log$(N/O) vs. 12\,+\,$\log$(O/H).  Different symbols represent 
different PNe samples (see legend).  Explanation of the data sets is given 
in the text.  
Red open circles (along with error bars) are the three Northern Spur PNe 
observed in Paper~I.  Representative error bars of our targets and those 
of \citet[][also \citealt{balick13} and \citealt{cor15}]{kwi12} are given 
in the lower-left corner. 
The symbols in Figures~\ref{Ne_vs_O}--\ref{S_vs_Ne_Ar} have the same 
meaning. 
}
\label{NO_vs_He_O}
\end{center}
\end{figure*}

\begin{figure*}
\begin{center}
\includegraphics[width=1.95\columnwidth,angle=0]{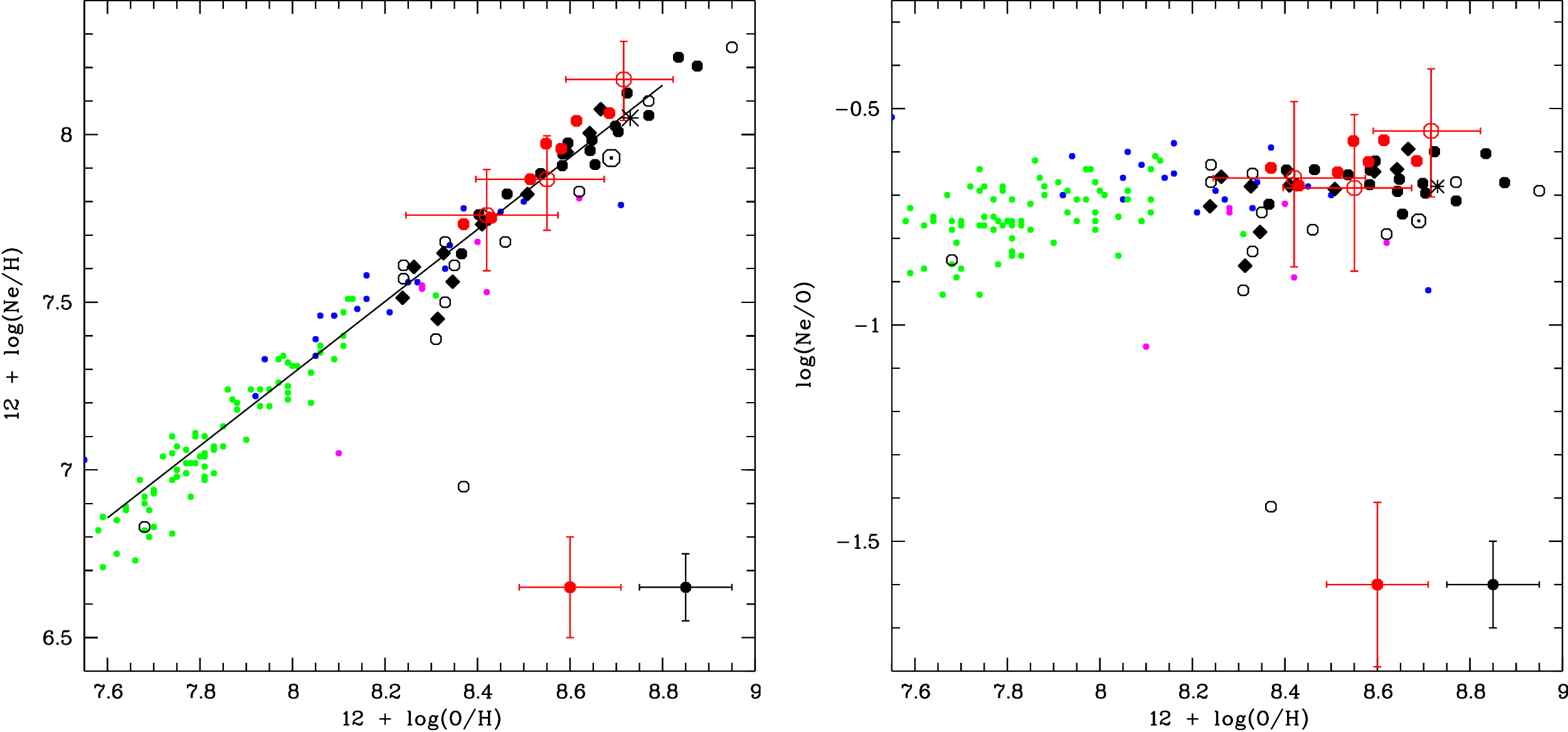}
\caption{Left: 12\,+\,$\log$(Ne/H) vs. 12\,+\,$\log$(O/H).  The solid black 
line is a least-squares linear fit to the H~{\sc ii} region data. 
Right: $\log$(Ne/O) vs. 12\,+\,$\log$(O/H).  The small colored (green, 
blue and magenta) dots are H~{\sc ii} regions and metal-poor blue compact 
galaxies (see references in the text).  The symbols in 
Figures~\ref{Ar_vs_O}--\ref{S_vs_Ne_Ar} have the same meaning.
}
\label{Ne_vs_O}
\end{center}
\end{figure*}

\begin{figure*}
\begin{center}
\includegraphics[width=1.95\columnwidth,angle=0]{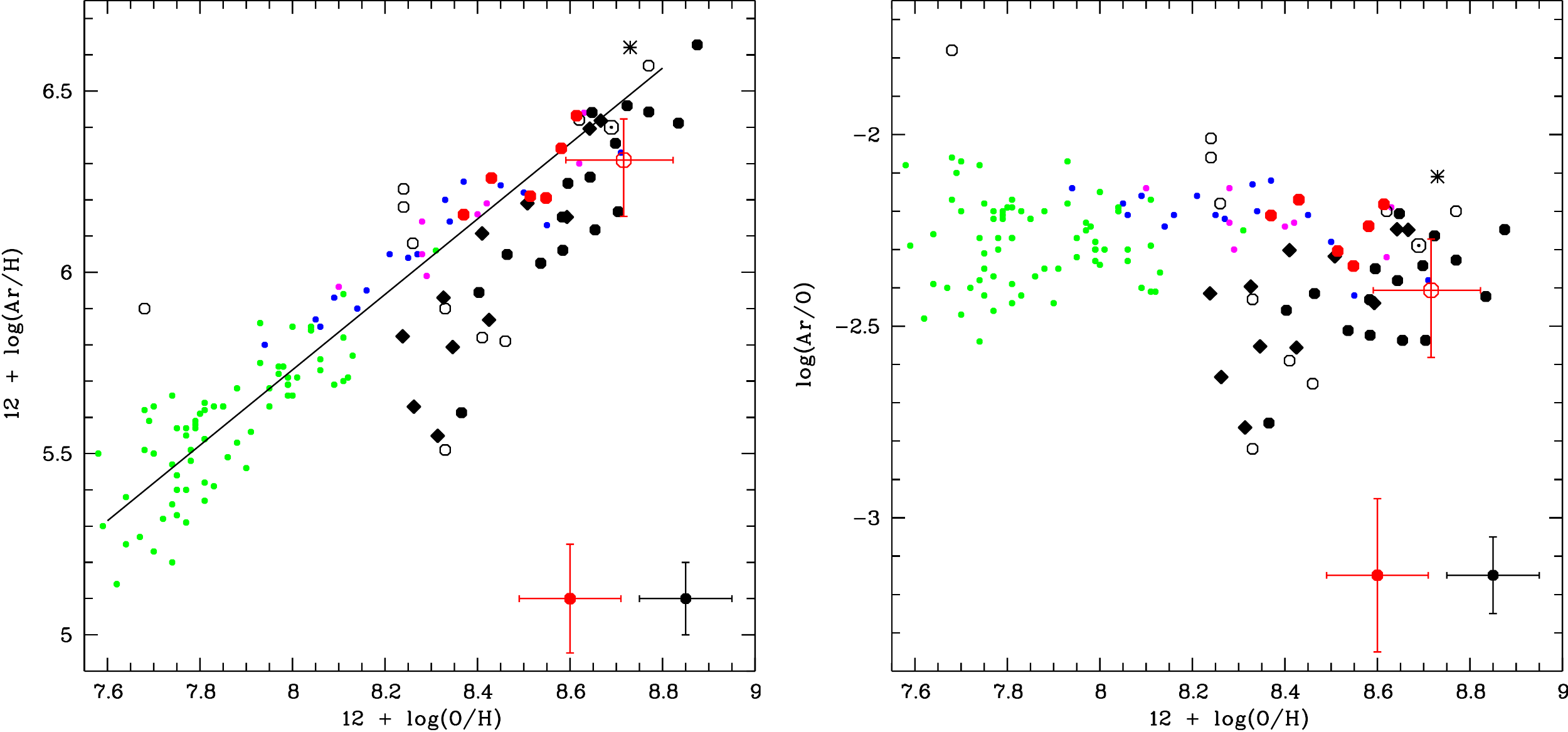}
\caption{Same as Figure~\ref{Ne_vs_O} but for 12\,+\,$\log$(Ar/H) vs. 
12\,+\,$\log$(O/H) (left) and $\log$(Ar/O) vs. 12\,+\,$\log$(O/H) (right). }
\label{Ar_vs_O}
\end{center}
\end{figure*}

\begin{figure*}
\begin{center}
\includegraphics[width=1.95\columnwidth,angle=0]{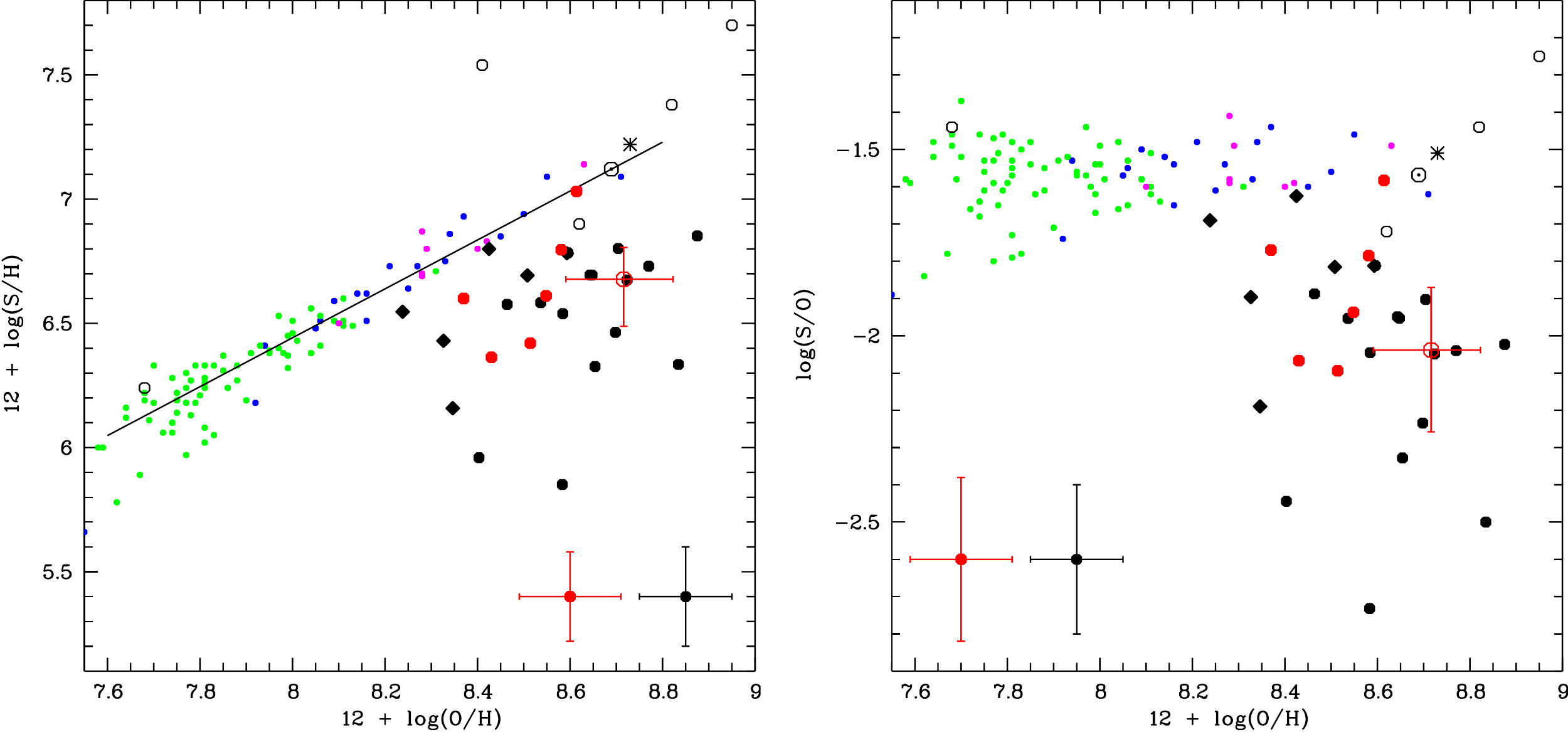}
\caption{Same as Figure~\ref{Ne_vs_O} but for 12\,+\,$\log$(S/H) vs.
12\,+\,$\log$(O/H) (left) and $\log$(S/O) vs. 12\,+\,$\log$(O/H) (right). }
\label{S_vs_O}
\end{center}
\end{figure*}

\begin{figure*}
\begin{center}
\includegraphics[width=1.95\columnwidth,angle=0]{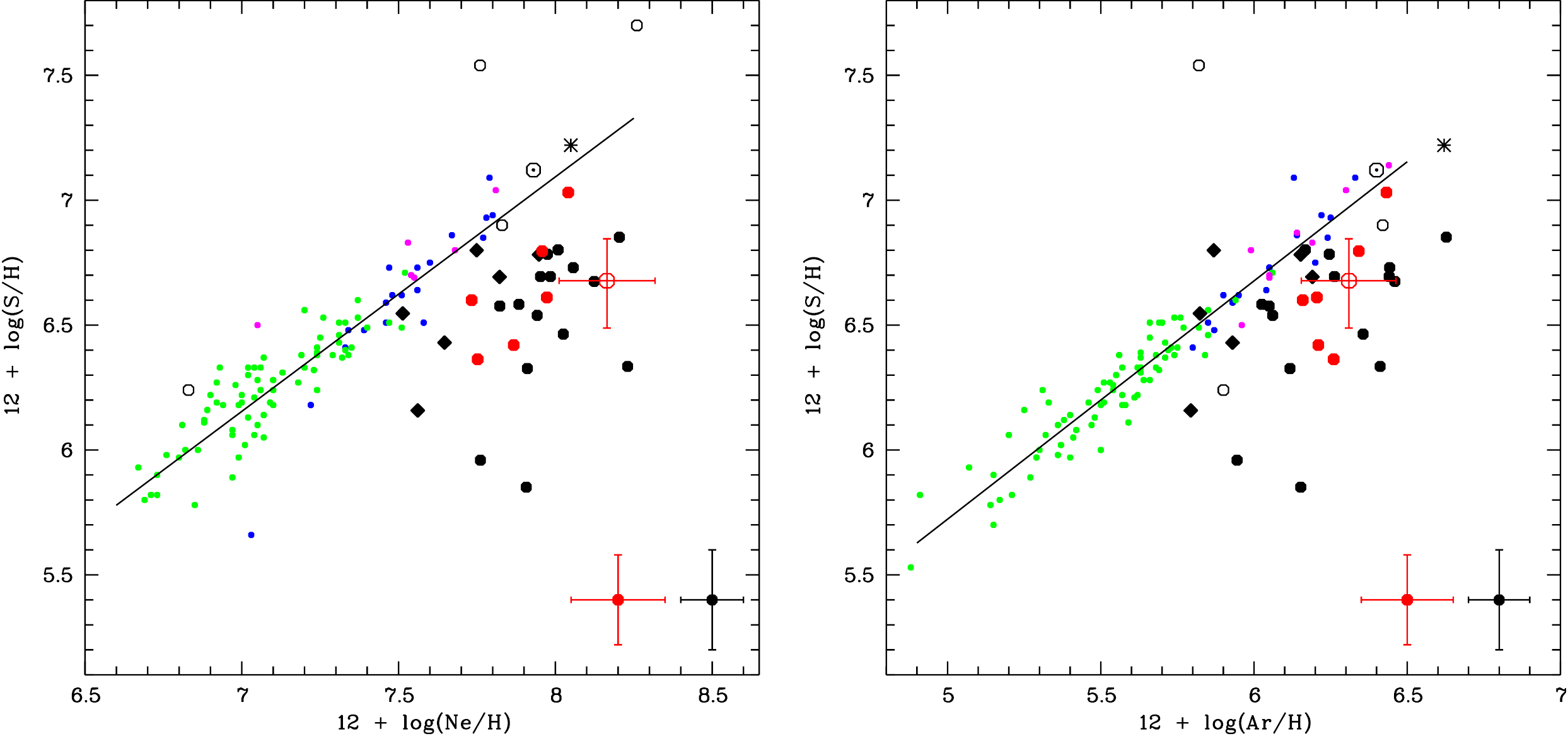}
\caption{Same as Figure~\ref{Ne_vs_O} but for 12\,+\,$\log$(S/H) vs.
12\,+\,$\log$(Ne/H) (left) and 12\,+\,$\log$(S/H) vs. 12\,+\,$\log$(Ar/H) 
(right). }
\label{S_vs_Ne_Ar}
\end{center}
\end{figure*}

Figures~\ref{NO_vs_He_O}--\ref{S_vs_Ne_Ar} show the abundance correlations 
of our sample as well as those for the M31 disk and bulge PNe from the 
literature.  Overplotted in these figures are the M31 outer-disk PNe 
observed by \citet{kwi12}, \citet{balick13} and \citet{cor15}, as well 
as the M31 disk and bulge sample observed by \citet{jc99}.  Also 
presented in the figures (except Figure~\ref{NO_vs_He_O}) are the 
extragalactic H~{\sc ii} region data, which set a baseline for the 
assumed tight linear behavior (in logarithmic scale) between different 
$\alpha$ elements:  The M101 H~{\sc ii} regions of \citet[][blue 
dots]{ken03}, the H~{\sc ii} regions in low-metallicity blue compact 
galaxies observed by \citet[][]{izo99} and \citet[][green dots]{izo12}, 
and the M31 H~{\sc ii} regions observed by \citet[][magenta dots]{zb12}. 
These plots also include the solar values \citep{asp09} and the Orion 
nebular abundances \citep{est04}.

Figure~\ref{NO_vs_He_O} shows the N/O vs. He/H (left) and N/O vs. O/H 
(right) correlations.  The Galactic disk PNe (mostly Type~I) observed by 
\citet{mil10} are also presented in this figure for purpose of comparison. 
The N/O and He/H ratios of our sample are generally lower than those of the 
Galactic Type~I PNe (Figure~\ref{NO_vs_He_O}, left).  Our PNe have N/O 
ratios much lower than 0.8, a criterion to distinguish the Galactic Type~I 
and II PNe \citep{kb94}.  The seven PNe in this study, together with the 
three Northern Spur PNe in Paper~I, are qualified to be classified as 
Type~II, i.e., their progenitor stars have relatively low masses.  The 
sample of \citet{mil10} shows a positive correlation between N/O and He/H 
and a decrease in N/O with oxygen, indicating that more massive progenitors 
produce nitrogen at the expense of oxygen.  Neither our targets nor the 
samples of \citet{kwi12}, \citet{balick13} and \citet{cor15} show such 
trends as clearly as the Galactic PNe. 


There is tight positive correlation between $\log$(Ne/H) and $\log$(O/H), 
as established based on the H~{\sc ii} region data \citep[e.g.,][]{mil10}. 
The neon-oxygen abundance distribution of our M31 PNe as well as those of 
\citet{kwi12}, \citet{balick13}, and \citet{cor15} generally agrees with 
this correlation within the errors, as shown in Figure~\ref{Ne_vs_O} 
(left).  Figure~\ref{Ne_vs_O} (right) shows the flatness of Ne/O in our 
sample.  Observations of the Galactic PNe have revealed that 
destruction/production of the $\alpha$ elements, particularly in the 
case of neon, does occur \citep{mil10}.  Figure~\ref{Ne_vs_O} shows 
that there is at lease one outlier in the sample of \citet{jc99}, 
indicating possible destruction of Ne.  However, the uncertainties of 
\citet{jc99} were not presented.  Argon of our sample is generally 
correlated with oxygen, following the pattern of H~{\sc ii} regions; 
however, the M31 outer-disk PNe of \citet{kwi12}, \citet{balick13}, and 
\citet{cor15} have systematically lower argon abundances 
(Figure~\ref{Ar_vs_O}). 
The exact cause of this systematic difference in the argon 
abundances is unclear, although the use of different ICFs could not 
be totally ruled out. 
The sample of \citet{jc99}, as well as the Type~I PNe of \citet{mil10}, 
have large scatter.

The sulfur anomaly, first noticed in Galactic PNe by \citet{hen04} and 
confirmed by \citet{mil10}, is also present in the M31 sample 
(Figure~\ref{S_vs_O}):  The sulfur abundances of the M31 PNe are 
significantly lower than those of H~{\sc ii} regions at similar oxygen 
abundance.  Besides, the scatter in the PN sulfur abundances for a given 
metallicity is much larger than that in H~{\sc ii} regions.  The sulfur 
anomaly is also evident in the diagrams of S/H vs. Ne/H 
(Figure~\ref{S_vs_Ne_Ar}, left) and S/H vs. Ar/H (Figure~\ref{S_vs_Ne_Ar}, 
right).  \citet{hen04} originally proposed that the apparently low sulfur 
abundances were likely due to ICF problems.  This argument was later 
confirmed by \citet{hen12}.  Under the physical conditions of PNe, a 
significant amount of sulfur is in the triply ionized stage S$^{3+}$, 
which can not be observed in the optical.  The presence of S$^{3+}$ is 
also indicated by the detection of He$^{2+}$ (e.g., the strong He~{\sc ii} 
$\lambda$4686 line of PN3), because the ionization potential of He$^{+}$ 
(54.4~eV) is higher than that of S$^{2+}$ (34.8~eV).  However, 
\citet[][also \citealt{hen12}]{mil10} found that the sulfur anomaly 
persists even after including the S$^{3+}$/H$^{+}$ abundances derived 
from the infrared {\it ISO} data.  It has been suggested that the sulfur 
anomaly could be due to the formation of dust such as MgS and FeS 
\citep{pb06}, but the exact amount of sulfur deficit caused by this 
mechanism is difficult to estimate.


\subsection{Stellar Population} \label{section4:c}

The He/H and N/O ratios as well as the abundance ratios of the 
$\alpha$-elements of our PNe generally span in the ranges within 
those of the disk PNe observed by \citet{kwi12}, \citet{balick13} and 
\citet{cor15}.  This indicates that our sample have probably similar 
properties as those of \citet{kwi12}.  Adopting the linear least-squares 
fit to the extinction parameter $c$(H$\beta$) versus PN core-mass 
relation presented by \citet[][Figure~4]{kwi12}, we estimate that the 
core masses of our sample are in the range $\sim$0.59--0.65\,$M_{\sun}$. 
The extinction versus core mass plot of \citet{kwi12} seems to 
indicate that this relation is insensitive to core mass below 
$\sim$0.59~$M_{\sun}$ when the extinction parameter $c$(H$\beta$) 
$\lesssim$0.20. 
Five object in our sample are in this range.  Here our extinction 
parameters $c$(H$\beta$) have been corrected for the Galactic average 
foreground extinction toward M31 \citep[$E$($B - V$) = 0.062;][]{sch98}, 
assuming the ratio $c$(H$\beta$)/$E$($B - V$) = 1.4 \citep{kl85}.  The 
initial masses of our target PNe are in the range 1.81--2.26\,$M_{\sun}$, 
as estimated using the initial-final mass relationship of white dwarfs 
given by \citet[][Equation~1]{cat08}. 
The progenitors' main-sequence lifetimes of our sample PNe are in the 
range 0.85--1.8\,Gyr, which was estimated using the stellar evolution 
model results of \citet[][Table~45]{sch92}.  Although the model tracks 
of \citet{vw94} for the post-AGB evolution were not used to constrain 
the core masses of our target PNe, the range of our core masses (also 
the initial masses) may be narrower than those of the sample studied 
by \citet{kwi12}.

Our estimate of the PN core masses is not robust, although the 
extinction versus PN core mass correlation presented in \citet{kwi12} 
can be physically explained, and has already been validated by 
\citet{jc99} and discussed in more detail in \citet{cj99}. 
This correlation is a simple consequence of the greater mass loss and 
faster evolution times of relatively high-mass stars \citep{cj99}. 
The mass-loss processes through the AGB wind are complex, and the 
circumstellar extinction depends largely on these ejection processes 
as well as on the central star evolution.  Thus the linear fit of 
\citet{kwi12} may still be crude, and it is inappropriate to simply 
use this relation to derive the core masses.  On this regard, our core 
masses are of large uncertainty.  With the future availability of deep 
GTC spectra obtained for more PNe in the substructures of M31\footnote{We 
have recently been awarded more observing time at the 10.4\,m GTC in 
Semester 2015B (September~1 2015 -- February~29 2016) to carry out deep 
spectroscopy of 11 more PNe in the substructures of M31 (GTC program 
\#GTC92-15B, PI: X.~Fang).}, we will construct detailed photoionization 
models for our PN sample, and more robust core masses will be derived. 
This will be presented in a separate paper.

Previous deep \emph{Hubble Space Telescope} Advanced Camera for Surveys 
(ACS) imaging revealed that the stellar fields at the Giant Stream are 
older than those on the disk \citep{ber15}, suggestive of an early-type 
progenitor.  The stream fields are generally consistent with the 
direction of target PN7 (Figure~\ref{m31_pne}).  The stellar field of 
the Northern Spur targeted by \citet{ber15} also well coincides with 
the position of target PN3 (Figure~\ref{m31_pne}).  In all ACS fields 
studied, \citet{ber15} found a significant burst of star formation 
2~Gyr ago, in rough agreement with the main-sequence ages we estimated 
for our target PNe.

\subsection{Radial Distribution of Oxygen} \label{section4:d}

\begin{figure*}
\begin{center}
\includegraphics[width=11cm,angle=-90]{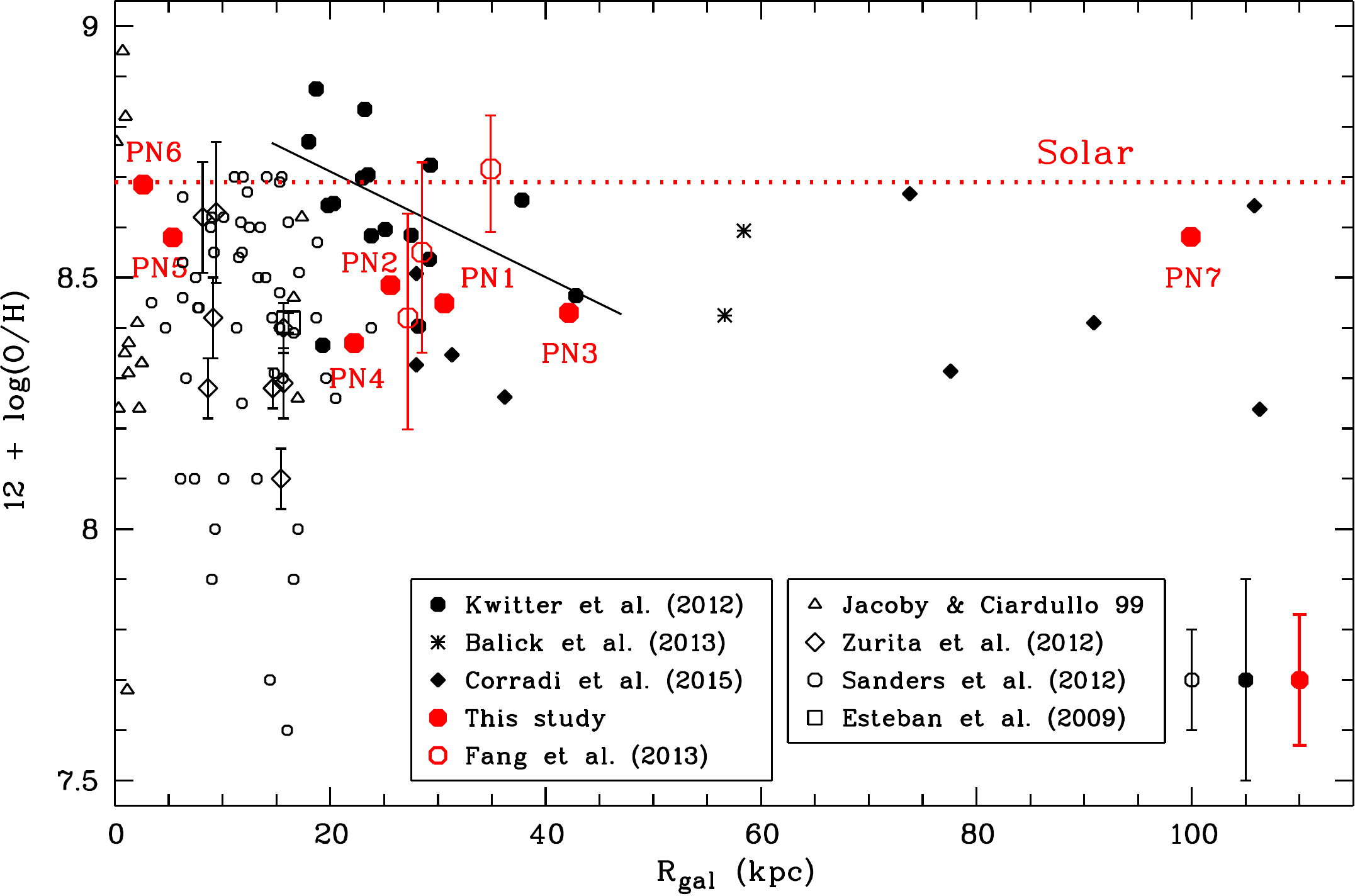}
\caption{
Distribution of oxygen abundances of PNe and H~{\sc ii} regions in M31. 
The black filled circles are the 16 M31 disk PNe observed by \citet{kwi12}; 
black asterisks are the two outer-disk PNe observed by \citet{balick13}; 
black filled diamonds are the nine outskirt PNe observed by \citet{cor15}. 
The black open circles are the PNe sample of \citet{san12}; black open 
triangles are the sample from \citet{jc99}.  Open diamonds are the nine 
disk H~{\sc ii} regions analyzed by \citet{zb12}.  The open square is an 
M31 H~{\sc ii} region (K932) studied by \citet{est09}.  Red open circles 
are the three Northern Spur PNe in Paper~I, and red filled circles are the 
seven PNe in this study.  Galactocentric distances have been rectified. 
The actual position of PN7 is $\sim$100\,kpc, as derived based on the 
three-dimensional structure of the Giant Stream \citep{mccon03}. 
Representative errors of our sample as well as those of \citet[][also 
\citealt{balick13} and \citealt{cor15}]{kwi12} and \citet{san12} are 
indicated in the lower right corner.  The black straight line is a linear 
fit to the disk sample of \citet{kwi12}.  The horizontal red dotted line 
is the solar oxygen abundance \citep{asp09}. }
\label{gradient}
\end{center}
\end{figure*}

The radial distribution of the oxygen abundances of our sample (the red 
open and filled circles) is presented in Figure~\ref{gradient}, where 
we also plot the M31 outer-disk/halo samples observed by \citet[][black 
filled circles]{kwi12}, \citet{balick13} and \citet[][black filled 
diamonds]{cor15}.  The PN samples of \citet[][black open circles]{san12} 
and \citet[][black open triangles]{jc99} are also plotted.  The M31 
H~{\sc ii} regions observed by \citet{zb12} and \citet{est09} are also 
presented in the figure for purpose of comparison.  Galactocentric 
distances (in kpc) have been rectified for the effects of projection 
on the plane of sky assuming that all objects are located on the disk, 
except PN7, which is located on the extension of the Giant Stream and 
whose distance to the center of M31 is probably $\geq$100~kpc if 
considering the three-dimensional structure of the Giant Stream 
\citep{mccon03}.  We assumed an inclination angle of 77\fdg7 for the M31 
disk and a position angle of 37\fdg7 for the M31 main axis \citep{dev58} 
in the projection rectification.

The M31 disk sample observed by \citet{kwi12} yield an oxygen gradient 
of $-$0.011$\pm$0.004~dex\,kpc$^{-1}$ [the target PN16, whose PN ID 
in \citet{mer06} is 1074, was not included in gradient fit because 
of absence of the [O~{\sc ii}] $\lambda$3727 line in the spectrum]. 
The two distant (halo) PNe observed by \citet{balick13} have the oxygen 
abundances comparable to that of the sun (Figure~\ref{gradient}), 
indicating a possible flattening in the outer disk.  Our sample, 
including the three Northern Spur PNe in Paper~I, spans a wide range in 
galactocentric distances and have rather homogeneous oxygen abundances. 
The oxygen abundance of PN7, the most distant PN in our sample, is 
consistent with those of the two outer-disk PNe of \citet{balick13} 
within the uncertainties.  However, they belong to different groups: 
PN7 is associated with the Giant Stream both spatially and kinematically, 
while the kinematics of Balick's sample follow the rotation pattern of 
the classical disk of M31 (Figure~\ref{orbit}).  PN5 and PN6 are close 
to the galaxy center and both have oxygen abundances close to that of 
PN7.  These three PNe are well associated with the extension of the Giant 
Stream of M31 (Figure~\ref{orbit}).

PN1, PN2 and PN3 are located in the Northern Spur and their oxygen 
abundances are consistent with those in Paper~I, which belong to the same 
substructure.  The average 12\,+\,$\log$(O/H) of our sample (10 PNe) is 
8.56$\pm$0.10, about 0.1~dex below the solar value \citep[8.69,][]{asp09}, 
while the mean value of the 18 outer-disk PNe of \citet{kwi12} and 
\citet{balick13} is 8.62$\pm$0.14, which is consistent with ours within 
dispersion.  The O/H ratios of our ten PNe are also generally consistent 
within the errors.  It is also worth noting that PN5 and PN6 have the 
oxygen abundances higher than the PN samples of \citet{jc99} and 
\citet{san12} at similar galactocentric distances, although the latter two 
both show large scatter.  The O/H ratios of these two PNe are also higher 
than those of the M31 H~{\sc ii} regions observed by \citet{est09} and 
\citet{zb12}.

\subsection{Possible Origin of the Substructure} \label{section4:e}

The carefully selected targets in this work, three belong to the 
Northern Spur and four associated with the Giant Stream substructure, 
have homogeneous oxygen abundances.  Although they can not be clearly 
distinguished from the outer-disk sample of \citet{balick13} by the 
abundances, their kinematics generally follow the stellar orbit that 
connects the Northern Spur and the Giant Stream.  The most intriguing 
object in our sample is PN7, a faint PN in the extension of the Giant 
Stream with a sky-projected angular distance of 3\fdg65.  This PN 
is located probably $>$100~kpc from the center of M31, according to the 
three-dimensional structure of Giant Stream \citep{mccon03}.  The 
similarity in its oxygen abundance with those of PN5 and PN6, which are 
both close to the bulge of M31 (Figure~\ref{gradient}), 
confirms that these three Stream PNe probably originated from the 
same stellar population. 
The spatial, kinematical and chemical information points to the 
possibility that these PNe have the same origin, and provides observational 
support for the stellar orbit, which was proposed by \citet{mer03} based 
on the kinematic study of $\sim$20 PNe spatially located on the M31 disk.

\citet{iba01b} initially suggested that either or both of M32 and NGC\,205
could be responsible for the Giant Stream, based on the geometrical 
alignment of the two satellites.  The panoramic survey of RGB stars in the 
M31 halo by \citet{fer02} reveals that M32 seems to be a more promising 
candidate for the origin of the stream.  These authors also inferred a 
possible connection between the Giant Stream and the extent of the Northern 
Spur, and even pointed out a strong warp of this connection orbit (if it 
does exist) near M31's nucleus.  Based on a kinematic study of $\sim$20 
disk PNe, \citet{mer03} constructed a model orbit that connects the Northern
Spur and the Giant Stream, and proposed that the two substructures are the 
tidal tails (one is the leading part, and the other the trailing part) of 
M32, given that this dwarf elliptical is close to the orbit both in 
position (projected on the $X$--$Y$ plane) and velocity.  \citet{mer03} 
also provided two extra arguments to support this hypothesis: 1) the RGB 
in the Northern Spur and the Giant Stream are both particularly red 
compared with the rest of the halo of M31 \citep{fer02}, and 2) M32 appears 
to be highly tidally stripped.  However, the exact position of M32 with 
respect to the disk of M31 remains yet to be determined \citep{mat98}.

The galaxy And~{\sc viii} is a tidally disrupted satellite of M31, 
and it was located very close to M32 in the $X$--$Y$ plane 
(Figure~\ref{orbit}) and might also be a candidate for the origin of 
the stellar stream \citep{mer03}.  This satellite was found to be 
strongly distorted, with a length of $\sim$10~kpc and a width of a few 
kiloparsecs, and contains 5--12 PNe and has a systematic velocity of 
$-$204~km\,s$^{-1}$ with respect to M31 \citep{mor03}.  And~{\sc viii} 
occupies the position where the Giant Stream meets the disk of M31, and 
the PNe in this satellite kinematically coincide with the model orbit 
(Figure~\ref{orbit}, bottom).  
However, the kinematics of And~{\sc viii} is opposite to that of M32, 
casting doubts on its possible origin of the stellar stream. 
Given the large scatter of the stellar stream both in positions and 
line-of-sight velocities, as discussed in \citet{mer03}, M32 is associated 
with the orbit.  Although spectroscopic observations of PNe in M32 has 
been carried out \citep{rsm99}, accurate abundance determination is 
scarce.  Future spectroscopic study of M32 PNe using the 8--10\,m class 
telescopes will help to confirm their true nature.

\section{Summary and Conclusion} \label{section5}

In order to investigate the possible origin of substructures of M31 using 
PNe as a tracer of chemical abundances, we carried out deep spectroscopy 
of PNe kinematically associated with the two most prominent substructures 
of M31, the Northern Spur and the Giant Stream.  Long-slit spectra were 
obtained for a carefully selected sample of seven PNe using the OSIRIS 
spectrograph attached to the 10.4\,m GTC in La Palma.  Careful data 
reduction was carried out to produce the highest quality 1D spectra ever 
taken for M31 PNe in the optical range ($\sim$3600--7800\,{\AA}).  The 
temperature-sensitive [O~{\sc iii}] $\lambda$4363 auroral line was well 
detected in the spectra of all targets and, in most cases, the [N~{\sc ii}] 
$\lambda$5755 auroral line was also detected, enabling determination of the 
electron temperature.  Electron densities were derived from the [Ar~{\sc 
iv}] and [S~{\sc ii}] nebular doublets.  We also estimated temperatures 
from the He~{\sc i} optical recombination line ratios, the first attempt 
of such effort for extragalactic PNe.  The He~{\sc i} temperatures are 
generally lower than those derived from the forbidden lines, consistent 
with observations in the Galactic PNe.

Ionic abundances were derived based on the [O~{\sc iii}] electron 
temperatures.  Elemental abundances of oxygen, nitrogen, neon, sulfur and 
argon were estimated using the ICF method.  The N/O ratios of our sample, 
including the three Northern Spur PNe in Paper~I, are in the range 
0.10--0.43, and the He/H ratios are $\sim$0.088--0.126, indicating that 
they are Type~II PNe, i.e., their progenitors probably have relatively 
low masses ($\leq$2~$M_{\sun}$).  
Using the empirical linear fit of the extinction versus core mass 
relation \citet{kwi12}, we roughly estimated that the core masses of 
our PN sample are in the range 0.59--0.65~$M_{\sun}$.  Using the stellar 
evolution model, we estimated the progenitors' main-sequence lifetimes 
of our PNe might be as old as $\lesssim$2~Gyr.  We reckon that our core 
masses as well as the main-sequence lifetimes are highly uncertain. 
With the availability of deep GTC spectra to be obtained for more PNe 
in the substructures of M31, we will carry out detailed photoionization 
modeling for our sample, and much more robust core masses will be 
derived. 
Our sample has an almost homogeneous distribution of the oxygen 
abundances.

Our GTC sample includes a target that is so far the most distant 
($\sim$3\fdg65) PN from the galaxy center ever discovered in M31.  It was 
newly captured and identified by the LAMOST spectroscopic survey, and is 
the first PN discovered in the outer streams of M31.  This PN is both 
spatially and kinematically related to the Giant Stream.  The other six 
PNe targeted by GTC are three PNe in the Northern Spur and three associated 
with the Giant Stream.  Our GTC run marks the first effort to carry out 
spectroscopic observations of PNe associated with the Giant Stream 
substructure.  All the ten substructure-associated targets (including the 
three Northern Spur PNe in Paper~I) are also well associated with the 
stellar orbit/streams that was proposed by \citet{mer03} to connect the 
Northern Spur and the Giant Stream, both spatially and kinematically.

The abundance-ratio distribution of our sample suggests that 
they belong to the same old population, while their spatial and 
kinematical distribution hints at the possibility that Northern Spur and 
the Giant Stream have the same origin, i.e., both of the two substructures 
are the debris of the tidal interaction between M31 and its satellite(s). 
This postulation agrees with an estimate of the old age for our targets, 
given that galactic interactions have long timescales. 
Judging from the spatial positions and kinematics of PNe therein, we 
emphasize the hypothesis of \citet{mer03} that the dwarf galaxy M32 
(and/or others) might be responsible for the stellar streams and thus 
the origin of the two substructures.  Deep spectroscopy of PNe in M32 
will help to assess this postulation.

This work, together with Paper~I, is a pioneering effort to study the 
substructures in M31 using PNe as tracers of chemistry.  Detailed 
spectroscopy using high-quality spectra obtained at a 10\,m-class 
telescope has proved to be very successful in deriving accurate abundance 
ratios, constraining stellar populations and thus the possible origin of 
substructures.  Effort as such may be extended to other Local Group 
galaxies. 

\section*{Acknowledgements}
This work is based on observations made with the Gran Telescopio Canarias 
(GTC), installed at the Spanish Observatorio del Roque de los Muchachos 
of the Instituto de Astrof\'{i}sica de Canarias, in the island of La 
Palma.  X.F. and M.A.G. acknowledge support from Spanish MICINN (Ministerio 
de Ciencia e Innovaci\'{o}n) grant AYA~2014-57280-P co-funded with 
FEDER funds.  R.G.B. acknowledges support from the MICINN AYA2010-15081 
grant.  Y.Z. thanks the Hong Kong General Research Fund (HKU7062/13P) 
for the financial support of this study.  The support and advice from the 
GTC staff, especially the GTC OSIRIS instrumental specialist, Antonio 
Cabrera-Lavers, is gratefully acknowledged.  We thank Enrique P\'{e}rez 
Jim\'{e}nez from the Instituto de Astrof\'{i}sica de Andaluc\'{i}a 
(IAA-CSIC) for comments and suggestions.  We also thank Michael R. 
Merrifield from the University of Nottingham for giving us the permission 
to reproduce the stellar orbit in Figure\,\ref{orbit} of this paper, 
based on Figure\,$2$ of \citet{mer03}.  We would also like to thank the 
anonymous referee whose comments have greatly improved the quality of 
this article.  This research has made use of NASA's Astrophysics Data 
System (http://adsabs.harvard.edu), the SDSS DR10 Science Archive Server 
(SAS; http://data.sdss3.org/) and the SIMBAD Astronomical Database 
(http://simbad.u-strasbg.fr/simbad/).
\\

{\it Facilities:} \facility{GTC (OSIRIS)}


\end{document}